\pdfoutput=1
\documentclass[12pt,letterpaper,titlepage,reqno,oneside]{article}
\usepackage[utf8]{inputenc}
\usepackage{graphicx}
\usepackage[super,comma]{natbib}
\usepackage[english]{babel}
\usepackage[figurename=FIG.,tablename=TABLE]{caption}
\usepackage{enumerate}
\usepackage{amsmath,amsfonts,amssymb,wasysym,latexsym,amsthm,bm}
\usepackage[mathscr]{eucal}
\usepackage{color}
\usepackage{subfig}
\usepackage{siunitx}
\usepackage{textcomp}
\usepackage{titlesec}
\usepackage{multirow}
\usepackage[subfigure,titles]{tocloft}
\usepackage{setspace}

\DeclareCaptionFormat{upper}{#1#2\uppercase{#3}\par}

\titleformat{\section}{\normalsize\bfseries}{\Roman{section}.}{1em}{\normalsize\uppercase}
\titleformat{\subsection}{\normalsize\bfseries}{\Alph{subsection}.}{1em}{}
\titleformat{\subsubsection}{\normalsize\itshape}{\arabic{subsubsection}.}{1em}{}

\renewcommand{\thesection}{\Roman{section}}

\cftpagenumbersoff{figure}
\cftpagenumbersoff{subfigure}

\makeatletter
\renewcommand\listoffigures{%
    \medskip\raggedright \textbf{LIST OF FIGURES}%
    \@mkboth{\MakeUppercase\listfigurename}%
        {\MakeUppercase\listfigurename}%
    \@starttoc{lof}%
}
\makeatother

\makeatletter
\def\p@subsection{\thesection.\,}
\makeatother

\begin{document}
\sisetup{detect-family,detect-display-math=true,exponent-product=\cdot,output-complex-root=\text{\ensuremath{i}},per-mode=symbol}

\author{%
Joris P. Oosterhuis\footnote{Author to whom correspondence should be addressed. E-mail: j.p.oosterhuis@utwente.nl}, Simon B\"uhler, and Theo H. van der Meer \\
\textit{Department of Thermal Engineering, University of Twente,}\\
\textit{Enschede, The Netherlands}\\ \\
Douglas Wilcox\\
\textit{Chart Inc., Troy, New York}%
}
\title{\LARGE{Jet pumps for thermoacoustic applications: design guidelines based on a numerical parameter study} \\
\textit{\large{Jet pump design guidelines}}
}
\date{\today}

\maketitle

\begin{abstract}
The oscillatory flow through tapered cylindrical tube sections (jet pumps) is characterized by a numerical parameter study. The shape of a jet pump results in asymmetric hydrodynamic end effects which cause a time-averaged pressure drop to occur under oscillatory flow conditions. Hence, jet pumps are used as streaming suppressors in closed-loop thermoacoustic devices. A two-dimensional axisymmetric computational fluid dynamics model is used to calculate the performance of a large number of conical jet pump geometries in terms of time-averaged pressure drop and acoustic power dissipation. The investigated geometrical parameters include the jet pump length, taper angle, waist diameter and waist curvature. In correspondence with previous work, four flow regimes are observed which characterize the jet pump performance and dimensionless parameters are introduced to  scale the performance of the various jet pump geometries. The simulation results are compared to an existing quasi--steady theory and it is shown that this theory is only applicable in a small operation region. Based on the scaling parameters, an optimum operation region is defined and design guidelines are proposed which can be directly used for future jet pump design. \\



\end{abstract}

\addtocounter{page}{2}

\section{Introduction}
\label{sec:intro}

One of the most promising configurations for thermoacoustic devices is the traveling wave configuration which generally consists of a closed-loop tube.\cite{Ceperley1979,Backhaus1999} This design however, has one disadvantage: the possibility of a time-averaged mass flow, or Gedeon streaming, to occur due to the looped geometry\cite{Gedeon1997}. This streaming leads to unwanted convective heat transport and reduces the efficiency of closed-loop thermoacoustic devices. 

\begin{figure}
\centering
\includegraphics[width=.5\textwidth]{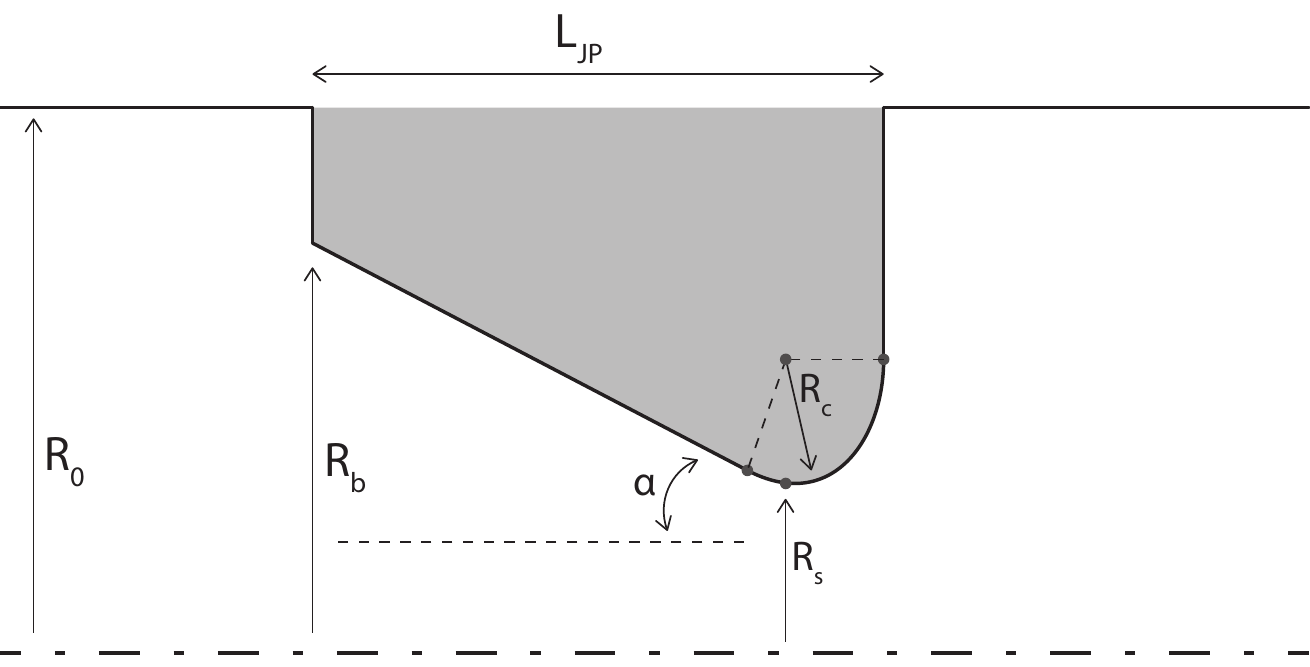}
\caption{Jet pump with parameters that define the geometry (not to scale). Bottom dashed line indicates center line, top solid line indicates outer tube wall. Reproduced from~\cite{Oosterhuis2015}.}
\label{fig:jetpumpgeom}
\end{figure}

A commonly used solution is the application of a jet pump.\cite{Backhaus2000,Swift1999,Biwa2007,Boluriaan2003} This is a tapered tube section which, due to the asymmetry in the hydrodynamic end effects, establishes a time-averaged pressure drop. By balancing the time-averaged pressure drop across the jet pump with that which exists across the regenerator of a thermoacoustic device, the Gedeon streaming can be canceled.\cite{Backhaus2000} Fig.~\ref{fig:jetpumpgeom} shows a schematic of a typical conical jet pump geometry with its corresponding dimensions. The two openings both have a different radius: $R_b$ for the big opening (left side) and $R_s$ for the small opening. Together with the jet pump length, $L_\mathit{JP}$, the jet pump taper angle $\alpha$ is defined. Furthermore, at the small opening a curvature $R_c$ is applied to increase the time-averaged pressure drop compared to a sharp contraction.
To estimate the performance of a jet pump, a quasi--steady model has been proposed by Backhaus and Swift.\cite{Backhaus2000} This model is based on minor losses in steady flow and assumes that the oscillatory flow can be approximated as two steady flows.\cite{Iguchi1982_u-shaped} Given the pressure drop generated by a pipe transition in steady flow,\cite{Idelchik2007}
\begin{equation}
\Delta p_\mathit{ml} = \frac{1}{2}K\rho u^2,
\end{equation}
and assuming a pure sinusoidal velocity in the jet pump, the time-averaged pressure drop across the jet pump can be estimated as
\begin{align}
\Delta p_{2,\mathit{JP}} & = \frac{1}{8}\rho_0 |u_{1,\mathit{JP}}|^2\left[(K_{\mathit{exp},s}-K_{\mathit{con},s}) + \left(\frac{A_s}{A_b}\right)^2(K_{\mathit{con},b}-K_{\mathit{exp},b})\right],
\label{eq:backhaus}
\end{align}
where $|u_{1,\mathit{JP}}|$ is the velocity amplitude at the small exit of the jet pump. The subscripts ``$s$'' and ``$b$'' indicate the small and big opening of the jet pump, respectively. $K_\mathit{exp}$ is the minor loss coefficient for abrupt expansion and $K_\mathit{con}$ is the minor loss coefficient for contraction, both are well documented for steady flows.\cite{Idelchik2007}

With the jet pump being an additional flow resistance, acoustic power will be dissipated due to viscous dissipation and vortex formation. Using the same quasi--steady approach, Backhaus and Swift derived an equation to estimate the amount of acoustic power dissipation associated with the jet pump as\cite{Backhaus2000}
\begin{align}
\Delta \dot{E}_\mathit{JP} & = \frac{\rho_0 |u_{1,\mathit{JP}}|^3A_s}{3\pi}\left[(K_{\mathit{exp},s}+K_{\mathit{con},s}) + \left(\frac{A_s}{A_b}\right)^2(K_{\mathit{con},b}+K_{\mathit{exp},b})\right].
\label{eq:backhaus_dE}
\end{align}

An optimal jet pump should establish the required amount of time-averaged pressure drop to cancel any Gedeon streaming with minimal acoustic power dissipation. This requires maximizing the difference in minor losses due to contraction and expansion while at the same time minimizing the sum of the minor loss coefficients.

Previous studies have shown however, that the accuracy of this quasi--steady approximation is limited and that there are other factors influencing the performance of a jet pump.\cite{Petculescu2003,Smith2003a,Oosterhuis2014,Oosterhuis2015,Tang2015} Petculescu and Wilen experimentally studied the influence of taper angle and curvature on minor loss coefficients.\cite{Petculescu2003} Good agreement was obtained between steady flow and oscillatory experiments for taper angles up to \SI{10}{\degree}. However, the minor loss coefficients determined were found to be strongly dependent on the taper angle used. A qualitative comparison between their findings and the current results will be provided in Section~\ref{sec:varAlpha_varRs}. Smith and Swift investigated a single diameter transition, corresponding to one end of a jet pump.\cite{Smith2003a} The measured pressure drop and acoustic power dissipation was found to be dependent on the dimensionless stroke length, the dimensionless curvature and the acoustic Reynolds number. Recently, Tang et al. investigated the performance of jet pumps numerically,\cite{Tang2015} but assumed a priori the quasi--steady approximation to be valid by modeling the flow as two separate steady flows. The negative effect of flow separation on the jet pump performance was identified which is in line with the current work.  

\begin{figure}
\centering
\subfloat[Oscillatory vortex pair on both sides and no jetting is observed.\label{fig:flowregime1}]{\includegraphics[width=.7\textwidth]{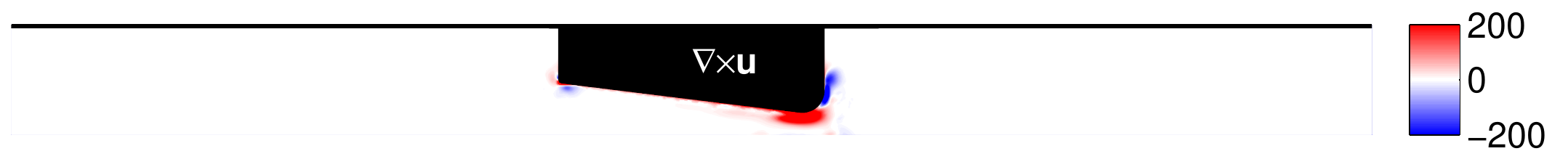}}\\
\subfloat[Propagating vortex to right side and oscillating vortex pair on left side of the jet pump.\label{fig:flowregime2}]{\includegraphics[width=.7\textwidth]{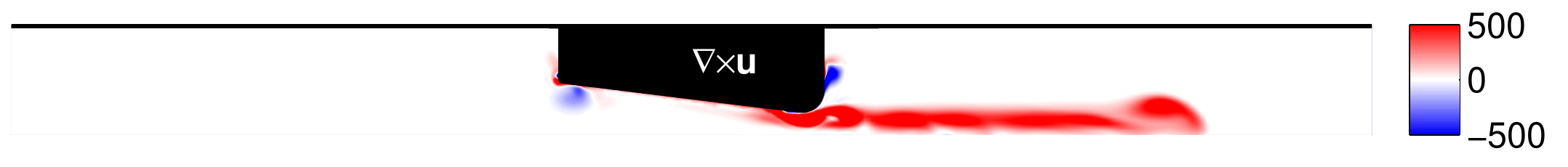}}\\
\subfloat[Propagating vortex on both sides of the jet pump. \label{fig:flowregime3}]{\includegraphics[width=.7\textwidth]{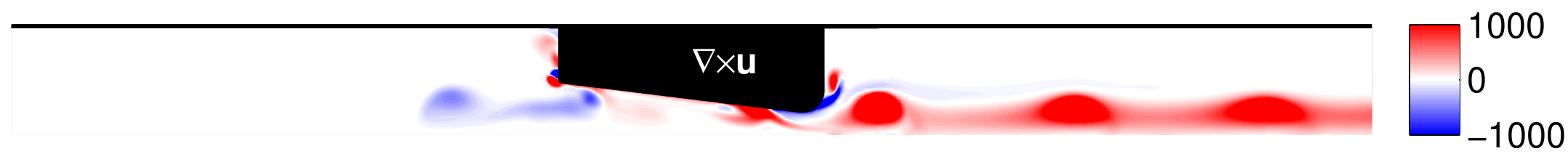}}\\
\subfloat[Vortices propagate fully through jet pump and flow separation inside the jet pump occurs. \label{fig:flowregime4}]{\includegraphics[width=.7\textwidth]{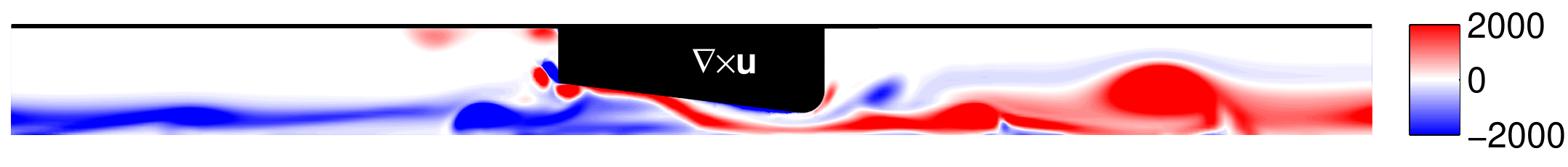}}
\caption{Four different flow regimes distinguished based on the instantaneous vorticity fields $\nabla \times \mathbf{u}$~$[\SI{1}{\per\s}]$ at $t=t_\mathit{max}$ around the jet pump for the $\alpha=\SI{7}{\degree}$ geometry (color online). Reproduced from~\cite{Oosterhuis2015}.}
\label{fig:flowregimes}
\end{figure}

Although some jet pump measurements are available in literature, a systematic parameter study that directly relates variations in wave amplitude and geometry to a jet pump's performance has, to the authors knowledge, not yet been addressed. A first step towards the investigation of a jet pump's performance in oscillatory flows has been made in previous work by scaling the jet pump geometry using two different Keulegan--Carpenter numbers and correlating this to the jet pump performance.\cite{Oosterhuis2015} Four different flow regimes were distinguished as a function of the wave amplitude. Fig.~\ref{fig:flowregimes} shows the observed vorticity fields in the vicinity of a jet pump at $t=t_\mathit{max}$ when $u_\mathit{JP}(t)>0$. In order of increasing wave amplitude these flow regimes can be described as follows. At low wave amplitudes, an oscillatory vortex pair exists on both sides of the jet pump but no vortex shedding is observed (Fig.~\ref{fig:flowregime1}). When the amplitude is increased, one-sided vortex propagation at the location of the small jet pump opening starts (Fig.~\ref{fig:flowregime2}). In this flow regime, vortex rings are shed from the jet pump waist to the right while on the big side of the jet pump an oscillatory vortex pair is still visible. Vortex shedding from the big side of the jet pump begins to occur at increased wave amplitudes (Fig.~\ref{fig:flowregime3}). Here vortex rings are shed from both jet pump openings directed outwards, but no interaction between the two sides is visible. Ultimately, a further increase in wave amplitude leads to interaction between the two jet pump openings (Fig.~\ref{fig:flowregime4}). Vortices being shed from the small jet pump opening now propagate in alternating directions and flow separation inside the jet pump occurs.

Albeit the described flow regimes were distinguished using the proposed scaling, the data set was too limited to determine whether the observed flow separation (Fig.~\ref{fig:flowregime4}) was geometrically initiated by an increase in the taper angle or by a decrease in the jet pump length. In this paper, a parameter study is performed using a computational fluid dynamics (CFD) model to further identify the relation between the four different flow regimes and the jet pump geometry. The influence of various geometric parameters including the jet pump taper angle, length, curvature and waist diameter on the jet pump performance is investigated. Based on the presented results, the existing scaling parameters are further extended. Furthermore, a comparison with the quasi--steady model is provided to determine under what conditions the approximation is applicable. Based on this parameter study, design guidelines for future jet pump design are proposed.

After a description of the used CFD model in Section~\ref{sec:modeling}, the various investigated jet pump geometries are introduced in Section~\ref{sec:geom_varAlpha_varRs}. The resulting flow regimes will be distinguished in Section~\ref{sec:flowregimes} and subsequently linked to the jet pump performance in Section~\ref{sec:varAlpha_varRs}. Finally, the influence of the jet pump curvature is investigated in Section~\ref{sec:var_Rc} before drawing final conclusions on the design guidelines in Section~\ref{sec:conclusions}.

\section{Modeling}
\label{sec:modeling}

A two-dimensional axisymmetric CFD model is developed using the commercial software package ANSYS CFX version 14.5,\cite{ANSYS2011} identical to the model used in previous work.\cite{Oosterhuis2015} In the following, the numerical model with the corresponding boundary conditions will be repeated briefly and the methods to derive the jet pump performance will be explained.

The different jet pump geometries are confined in an outer tube with radius $R_0=\SI{30}{\mm}$. The length of the outer tube on either side of the jet pump is $L_0=\SI{500}{\mm}$ when $f=\SI{100}{\Hz}$. This length is scaled relative to the wavelength to avoid the jet pump being placed at a velocity node for simulations performed at other frequencies. In all cases, air at a mean pressure of $p_0=\SI{1}{atm}$ and a mean temperature of $T_0=\SI{300}{\kelvin}$ is used as the working fluid.

\subsection{Numerical model}

The unsteady, fully compressible Navier-Stokes equations are discretized using a high resolution advection scheme in space and a second order backward Euler scheme in time.\cite{ANSYS2012_NavierStokes} The system of equations is closed using the ideal gas law as the equation of state and the energy transport is calculated using the total energy equation, including viscous work terms. Based on the critical Reynolds number defined in Section~\ref{sec:results}, all presented results fall within the laminar regime so no additional turbulence modeling is applied. The time-step $\Delta t$ is chosen such that each wave period is discretized using 1000 time-steps. A total of ten wave periods are simulated to achieve a steady periodic solution and the last five wave periods are used for further analysis.

The acoustic wave is generated on the left boundary of the domain using a sinusoidal velocity boundary condition with a specified frequency and velocity amplitude. To control the wave propagation over the right boundary of the computational domain, a dedicated time-domain impedance boundary condition is developed and implemented in ANSYS CFX. The approach is based on the work of Polifke et al. and allows the application of a complex reflection coefficient at the boundary.\cite{Huber2008,Kaess2008} A detailed validation and explanation of the implementation can be found in the work of Van der Poel, which was carried out as part of the current research.\cite{VanderPoel2013} In all cases described here, a reflection coefficient of $|R|=0$ is specified to simulate a traveling wave on the right side of the jet pump. The combination of the velocity boundary condition and the time-domain impedance boundary condition results in a time-averaged volume flow that is on average less than \SI{0.5}{\percent} of the acoustic volume flow rate.

On the radial boundary of the outer tube (at $r=R_0$), a slip adiabatic wall boundary condition is imposed as the pipe losses in this part of the domain are currently not of interest. To correctly simulate the minor losses in the jet pump, a no-slip adiabatic wall boundary condition is used at the walls of the jet pump.

The choice for a two-dimensional axisymmetric model to simulate flow separation might need some additional explanation as in planar diffusers the flow separation can be asymmetric.\cite{Cherry2008,Tsui1995} A clear distinction should be made between flow separation in planar and conical diffusers. The steady flow separation in conical diffusers has been investigated previously and no visible asymmetry of the flow was reported under laminar conditions.\cite{Sparrow2009} Furthermore, the oscillatory nature of the flow prevents the flow from developing asymmetric instabilities, even in planar diffuser geometries.\cite{Nabavi2010,King2011} These observations motivate the applicability of a two-dimensional axisymmetric model for the current situation of pure oscillatory flow through a conical geometry.

\subsection{Computational mesh}

The used spatial discretization is validated in previous work and will be only briefly described here.\cite{Oosterhuis2015} The computational mesh consists of an unstructured part within \SI{50}{\mm} from the jet pump and a structured mesh in the rest of the domain. In both parts quadrilateral elements are used. In the jet pump region, a maximum element size of \SI{1}{\mm} is used which is refined up to \SI{0.5}{\mm} near the jet pump waist. Moreover, to be able to accurately resolve the flow separation, a refinement in the viscous boundary layer is applied such that a minimum of 10~elements are located within one viscous penetration depth distance $\delta_\nu$ from the jet pump wall. For each simulated wave frequency, the mesh distribution is adjusted according to the viscous penetration depth, $\delta_\nu = \sqrt{2\mu/\omega\rho}$. In the quadrilateral part of the mesh, a fixed radial mesh resolution of \SI{1}{\mm} is used whereas the mesh size in the axial direction grows from \SI{1}{\mm} at a distance of \SI{50}{\mm} from the jet pump up to \SI{10}{\mm} at the extremities of the domain.

In additional to the mesh validation performed in previous work,\cite{Oosterhuis2015} the accuracy of the employed computational mesh has been investigated specifically for the prediction of flow separation. The jet pump geometry with a taper angle of $\alpha=\SI{15}{\degree}$ (No.~4 in Table~\ref{tab:geom_varalpha}) has been used at a frequency of \SI{100}{\Hz} and at a wave amplitude where flow separation is expected ($\xi_1/D_s\cdot\alpha=0.85$, Eq.~\ref{eq:x1DsAlpha}). The mesh size in the jet pump region is in two steps refined from \SI{1}{\mm} to \SIlist{0.5;0.25}{\mm}. At the same time, the mesh size in the jet pump waist region is refined from \SI{0.5}{\mm} to \SIlist{0.2;0.1}{\mm}. Additionally, the effect of a mesh refinement in the viscous boundary layer is investigated by increasing the number of elements from 10 to 20. The difference in the dimensionless pressure drop (Eq.~\ref{eq:dp2_star}) and dimensionless acoustic power dissipation (Eq.~\ref{eq:dE2_star}) between the investigated meshes is less than 0.04 and 0.05, respectively. Furthermore, both the time and location where the flow first separates do not vary more than $\Delta t/T=\SI{8e-3}{}$ and $\Delta x=\SI{0.4}{\mm}$ between the different mesh resolutions.

\subsection{Data analysis}

In order to determine the jet pump performance from the transient CFD solution, some additional analysis steps are performed. The first order amplitudes of all physical quantities are calculated by a discrete Fourier transformation for the specified wave frequency using the solution from the last five simulated wave periods. To obtain the time-averaged quantities, the CFD solution is averaged over the last five wave periods in order to eliminate the first order contribution. The streaming velocity is calculated using a density-weighted time-average: $\mathbf{u}_2 = \langle\rho\mathbf{u}\rangle/\langle\rho\rangle$. The velocity amplitude in the jet pump waist, $|u_{1,\mathit{JP}}|$, is calculated using an area-weighted average over the cross-section. Subsequently, the local acoustic displacement amplitude in the jet pump waist $\xi_1$ can be calculated under the assumption of a sinusoidal jet pump velocity:
\begin{equation}
\xi_1 = \frac{|u_{1,\mathit{JP}}|}{2 \pi f}.
\label{eq:x1JP} 
\end{equation}
To scale the displacement amplitude to the jet pump geometry, a Keulegan--Carpenter number is defined based on the jet pump waist diameter:
\begin{equation}
KC_D = \frac{\xi_1}{D_s}, \label{eq:KCd}
\end{equation}
where $D_s = 2R_s$ is the jet pump waist diameter. This $KC_D$ is also referred to as a ``dimensionless stroke length'' or ``dimensionless displacement amplitude'' and can be rewritten to $Re_D/S^2$ where $Re_D$ is an acoustic Reynolds number based on diameter and $S=\sqrt{\omega D^2/\nu}$ the Stokes number~\cite{Smith2003a,King2011,Holman2005}.

\subsection{Jet pump geometries}
\label{sec:geom_varAlpha_varRs}

The described computational model is used to investigate a large variety of conical jet pump geometries and wave amplitudes. By varying either the jet pump length, the size of one of the two openings or the local curvature at the jet pump waist, 19 different geometries are considered. The different investigated geometries are all variations based on one reference geometry (denoted by ``\textit{ref}'' in all tables). The reference geometry is a jet pump with a \SI{7}{\degree} taper angle, all other reference dimensions are denoted in Table~\ref{tab:geom_ref}.
\begin{table}
\centering
	\caption{Dimensions of the reference jet pump geometry.}
	\label{tab:geom_ref}
	\begin{tabular}{ll}
	$R_0$ & \SI{30}{\mm}\\
	$R_b$ & \SI{15}{\mm}\\
	$R_s$ & \SI{7}{\mm}\\
	$R_c$ & \SI{5}{\mm}\\
	$L_\mathit{JP}$ & \SI{70.5}{\mm}\\
	$\alpha$ & \SI{7}{\degree}\\
	\end{tabular}
\end{table}

Varying the jet pump taper angle while keeping the waist diameter constant can be achieved by either changing the jet pump length $L_\mathit{JP}$ or changing the radius of the big opening $R_b$. Because the jet pump length is not taken into account in the quasi--steady approximation (Eq.~\ref{eq:backhaus}), the predicted jet pump performance is independent of the jet pump length. In a similar way, the radius of the big opening is only a weak parameter in the quasi--steady approximation as long as $R_b \gg R_s$. Both approaches are considered here to see what actually influences the jet pump performance: the distance between the two openings ($L_\mathit{JP}$) or the taper angle of the jet pump inner surface ($\alpha$). Table~\ref{tab:geom_varalpha} shows the dimensions of the investigated geometries where the taper angle is changed with respect to the reference case by either varying the jet pump length (cases 1-5) or by varying the big radius (cases 6-9). The jet pump waist radius and curvature are kept constant at $R_s=\SI{7}{\mm}$ and $R_c=\SI{5}{\mm}$, respectively. This results in a dimensionless curvature of $\chi = R_c/D_s=0.36$ which is considered a ``smooth'' contraction.\cite{Idelchik2007} Correspondingly, the contraction minor loss coefficient at the jet pump waist ($K_\mathit{con,s}$ in Eq.~\ref{eq:backhaus}) is expected to be negligible which will enhance the overall time-averaged pressure drop.
\begin{table}
\centering
	\caption{Dimensions of jet pump geometries with varied taper angles $\alpha$ by either changing the jet pump length or the big radius. Constant radius of the jet pump waist $R_s=\SI{7.0}{\mm}$ and constant curvature of the small opening $R_c=\SI{5.0}{\mm}$.}
	\label{tab:geom_varalpha}
	\begin{tabular}{lclll}
	No. & symbol & $\alpha$ & $L_\mathit{JP}$ & $R_b$\\
	\hline
	1 & {\Large$\bullet$} & \SI{3}{\degree} & \SI{157.8}{\mm} & \SI{15.0}{\mm}\\
	2 & $\blacksquare$ & \SI{5}{\degree} & \SI{96.7}{\mm} &  \SI{15.0}{\mm}\\
	\textit{ref} & $\times$ & \textit{\SI{7}{\degree}} & \textit{\SI{70.5}{\mm} }&  \SI{15.0}{\mm}\\
	3 & $\blacktriangledown$ & \SI{10}{\degree} & \SI{50.8}{\mm} & \SI{15.0}{\mm}\\
	4 & $\blacklozenge$ & \SI{15}{\degree} & \SI{35.5}{\mm} &  \SI{15.0}{\mm}\\
	5 & $\blacktriangleright$ & \SI{20}{\degree} & \SI{27.9}{\mm} & \SI{15.0}{\mm}\\
	\hline
	6 & {\Large$\bullet$} &\SI{3}{\degree} & \SI{70.5}{\mm} & \SI{10.4}{\mm}\\
	7 & $\blacksquare$ &\SI{5}{\degree} & \SI{70.5}{\mm} & \SI{12.7}{\mm}\\
	8 & $\blacktriangledown$ &\SI{10}{\degree} & \SI{70.5}{\mm} & \SI{18.5}{\mm}\\
	9 & $\blacklozenge$ & \SI{15}{\degree} & \SI{70.5}{\mm} & \SI{24.4}{\mm}\\
	\end{tabular}
\end{table}

In a different set of simulations, both the taper angle and the jet pump length are fixed while the jet pump waist diameter is changed (cases 10-13). Furthermore, the jet pump curvature is adjusted to maintain the same dimensionless curvature. The dimensions of this set of geometries are shown in Table~\ref{tab:geom_varRs}.
\begin{table}
\centering
	\caption{Dimensions of jet pump geometries with varied waist radius $R_s$. Constant taper angle $\alpha=\SI{7}{\degree}$ and jet pump length $L_\mathit{JP}=\SI{70.5}{\mm}$.}
	\label{tab:geom_varRs}
	\begin{tabular}{llll}
	No. & $R_s$ & $R_b$ & $R_c$\\
	\hline
	10 & \SI{3.0}{\mm} & \SI{11.4}{\mm} & \SI{2.1}{\mm} \\
	11 & \SI{5.0}{\mm} & \SI{13.2}{\mm} & \SI{3.6}{\mm} \\
	\textit{ref} & \textit{\SI{7.0}{\mm}} & \textit{\SI{15.0}{\mm}} & \textit{\SI{5.0}{\mm} }\\
	12 & \SI{9.0}{\mm} & \SI{16.8}{\mm} & \SI{6.4}{\mm} \\
	13 & \SI{11.0}{\mm} & \SI{18.6}{\mm} & \SI{7.9}{\mm} \\
	\end{tabular}
\end{table}

In the following, the observed flow regimes and jet pump performance will be discussed for the cases where either the jet pump length or the size of one of the two openings is varied. In Section~\ref{sec:var_Rc}, the effect of the jet pump waist curvature will be investigated.

\section{Results and discussion}
\label{sec:results}

For each geometry, various wave amplitudes and consequently, various displacement amplitudes are simulated. The wave amplitude is varied by specifying different values of the velocity amplitude at the left boundary condition. It is important to ensure that the simulated wave amplitudes fall within the laminar regime as URANS turbulence models, especially in combination with wall functions, are known to lose their validity for relaminarizing and separating flows which is an important part of the current data set.\cite{Cherry2008,Fugal2005,Michelassi1993} Other approaches to model turbulence such as large eddy simulation (LES) or direct numerical simulation (DNS) are currently not feasible for a parameter study of the size presented here. In order to determine what wave amplitudes fall within the laminar regime, the critical Reynolds number derived by Ohmi and Iguchi for oscillatory pipe flow is used,\cite{Ohmi1982a,Antao2013}
\begin{equation}
Re_c = 305\left(\frac{D}{\delta_\nu}\right)^\frac{1}{7},
\label{eq:Rec}
\end{equation}
where the Reynolds number is defined based on the viscous penetration depth: $Re = |u_1|\delta_\nu\rho_0/\mu_0$. Normally, the velocity amplitude is directly proportional to the frequency for a given displacement amplitude, but the velocity amplitude where a transition to turbulence occurs ($Re=Re_c$) scales with $f^{4/7}$. From previous work, it was concluded that the jet pump performance scales with the acoustic displacement amplitude instead of the jet pump velocity amplitude.\cite{Oosterhuis2015} Consequently, by lowering the driving frequency all displacement amplitudes of interest are investigated. The majority of the presented simulations are carried out with $f=\SI{100}{\Hz}$ while some additional simulations are using frequencies of \SIlist{10;20;30;50;200}{\Hz}. All data presented in this paper represents a total of 197 simulations having a total single core computational time of 391 days on an Intel Core i7 CPU. 

From the simulated jet pump performance, a scaling parameter is introduced which best aligns the jet pump performance for the various investigated geometries and correctly incorporates the effect of the jet pump taper angle:
\begin{equation}
\frac{\xi_1}{D_s}\cdot\alpha.
\label{eq:x1DsAlpha}
\end{equation}
This scaling parameter is essentially the Keulegan--Carpenter number (Eq.~\ref{eq:KCd}) multiplied by the taper angle (in radians). 

\subsection{Flow regimes}
\label{sec:flowregimes}

\begin{figure}
\centering
\subfloat[All four flow regimes. Dashed lines determine the bounds of the different flow regimes.\label{fig:flowregimes_KCd_x1DsAlpha}]{\includegraphics[width=.5\textwidth]{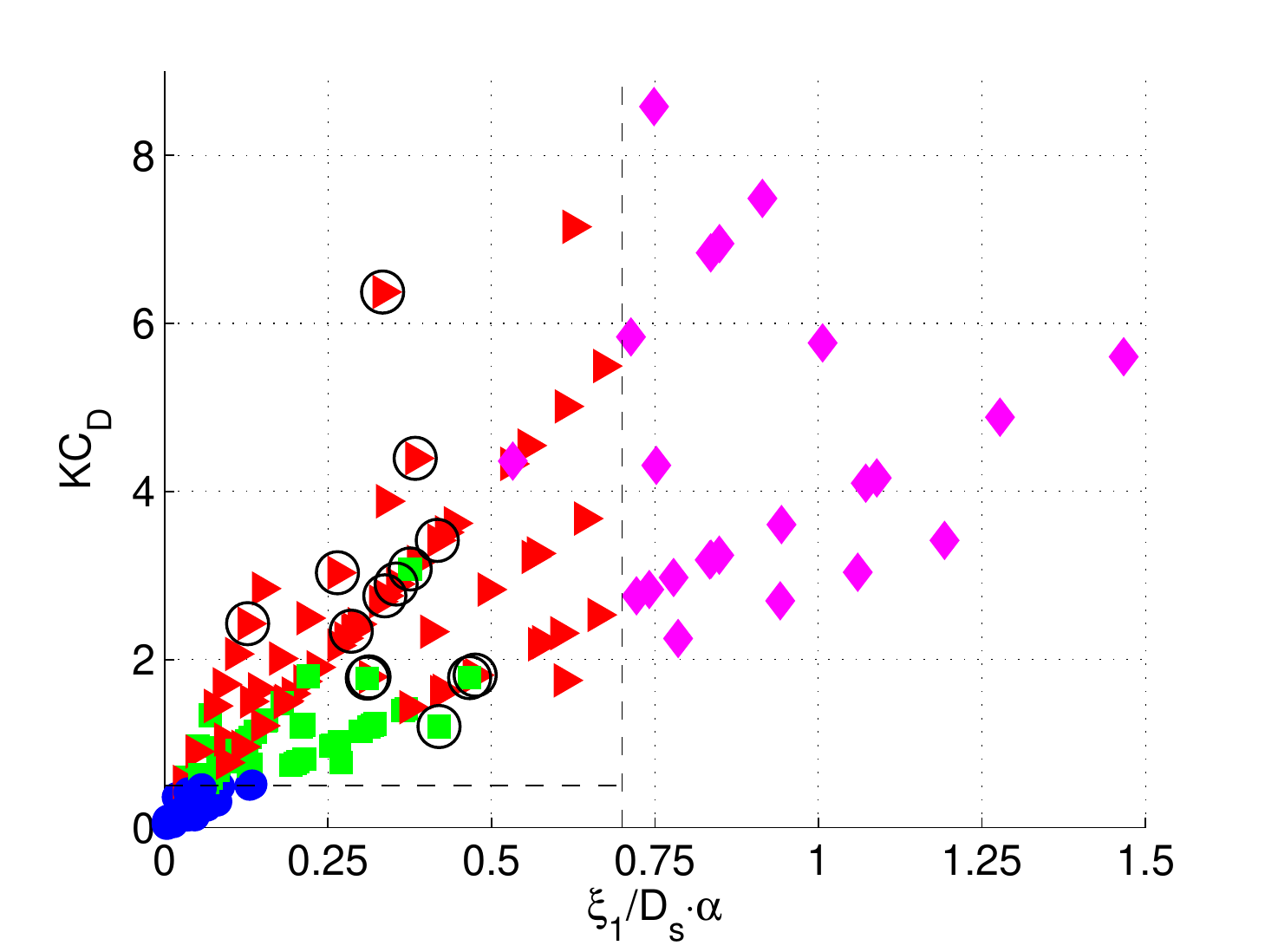}}
\subfloat[Flow regimes with one-sided ($\blacksquare$) and two-sided ($\blacktriangleright$) vortex propagation. The other two flow regimes are not shown. The local Keulegan--Carpenter number at the big jet pump opening, $KC_{D,b}$, is on the vertical axis. The flow regimes are separated by horizontal dashed line at $KC_{D,b}=0.15$.\label{fig:flowregimes_KCdb_x1Dsalpha}]{\includegraphics[width=.5\textwidth]{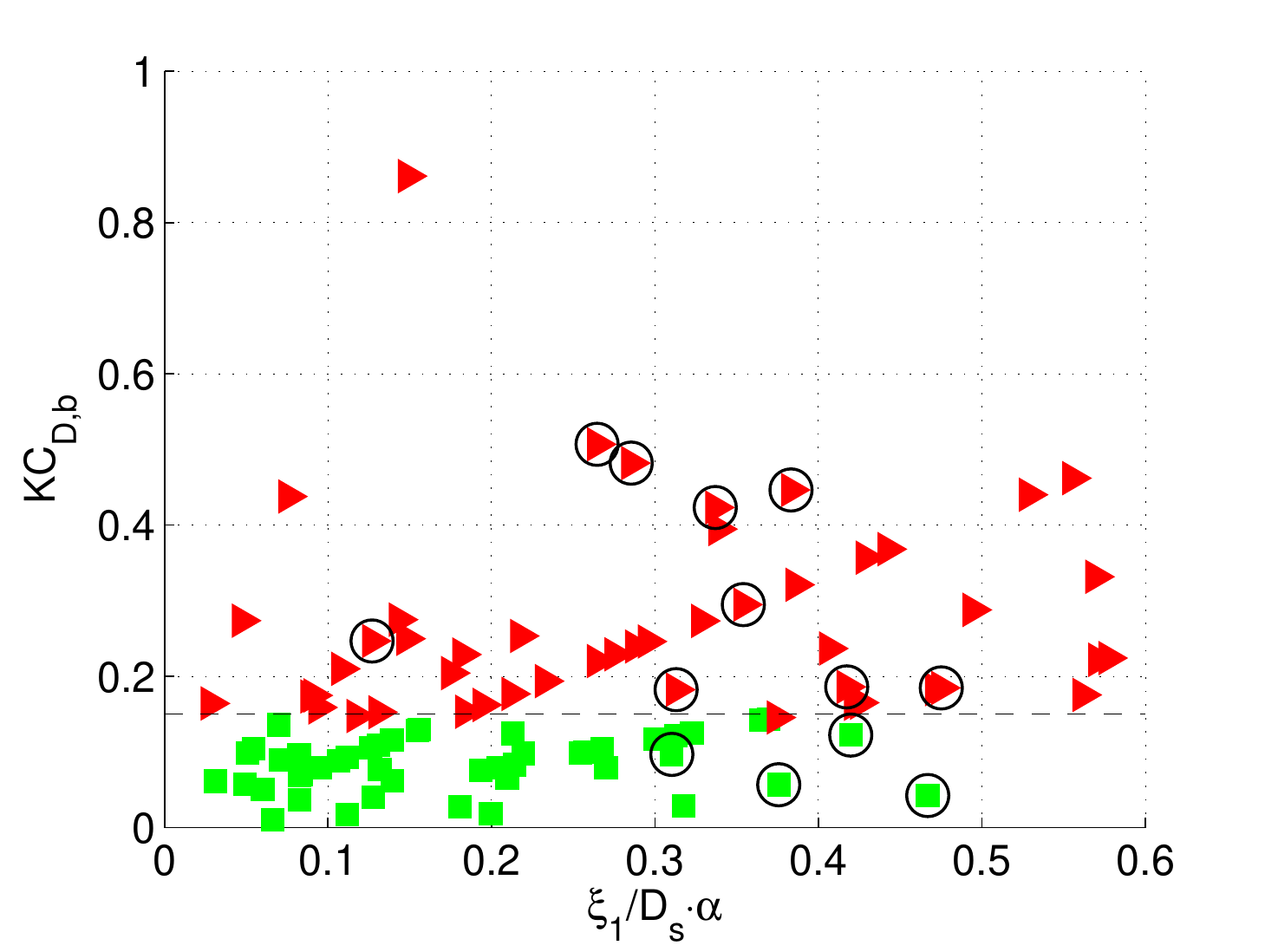}}
\caption{Observed flow regimes using jet pump geometries with varied taper angle and waist diameter plotted in two variable spaces. The different flow regimes are distinguished using different symbols: {\Large$\bullet$} for oscillatory vortices, $\blacksquare$ for right-side vortex propagation only, $\blacktriangleright$ for two-sided vortex propagation and $\blacklozenge$ for flow separation. The dashed lines represent the determined bounds of the flow regimes and the open circles indicate the points where the maximum $\Delta p_2^*$ was found for a specific jet pump geometry (color online).}
\label{fig:flowregimes_varAlpha+varRs}
\end{figure}

Because the jet pump performance is found to be a strong function of the flow regime, the observed flow regimes will first be discussed. By studying the transient vorticity field and streaming velocity field of all simulations, a differentiation is made between the four flow regimes that have been illustrated in Fig.~\ref{fig:flowregimes}. The flow regimes that have been observed using various wave amplitudes and jet pump geometries are all presented in the ($\xi_1/D_s\cdot\alpha, KC_D$) space in Fig.~\ref{fig:flowregimes_KCd_x1DsAlpha}. As such, the horizontal axis includes the taper angle while the vertical axis represents the Keulegan--Carpenter number without the contribution of the taper angle. The different symbols represent the different flow regimes: circles for oscillatory vortex pairs, squares for right-sided vortex propagation, triangles for two-sided vortex propagation and diamonds for flow separation and alternating vortex shedding. The open circles indicate the position where the maximum dimensionless pressure drop was achieved with a specific jet pump geometry.

Two flow regimes are easily distinguished in Fig.~\ref{fig:flowregimes_KCd_x1DsAlpha}. For small values of the Keulegan--Carpenter number, $KC_D<0.5$, no vortex shedding is observed. This is indicated by the horizontal dashed line in Fig.~\ref{fig:flowregimes_KCd_x1DsAlpha}. In this regime, no asymmetry in minor losses occurs and consequently the time-averaged pressure drop is zero. The transition from this flow regime to the situation where vortex shedding is first observed is independent of the taper angle and occurs when $KC_D>0.5$. Above this value the flow field becomes asymmetric resulting in a positive time-averaged pressure drop. This is in line with the work of Holman et al. on synthetic jets where a formation criterion for vortex shedding is derived of the form $KC_D=Re_D/S^2>C$ where $C$ is a constant depending on the geometry of the orifice.\cite{Holman2005}

In Fig.~\ref{fig:flowregimes_KCd_x1DsAlpha} the flow separation regime (diamonds) can be distinguished from the other three flow regimes and occurs when $\xi_1/D_s\cdot\alpha>0.7$ (dashed vertical line). Because this transition is clearly determined by the scaling parameter as introduced in Eq.~\ref{eq:x1DsAlpha}, the taper angle plays an important role in the initiation of the flow separation regime. This flow regime is dominated by vortices being shed from the jet pump waist in alternating directions. This results in the flow field again becoming more and more symmetric. Despite the interesting flow phenomena occurring in this flow regime,\cite{Oosterhuis2015} it is of little practical interest when designing jet pumps as the achieved jet pump effectiveness is low. Consequently, the area of interest for jet pump applications is bounded to $KC_D>0.5$ and $\xi_1/D_s\cdot\alpha<0.7$.

Inside this area, two flow regimes occur which cannot solely be separated based on $KC_D$ or $\xi_1/D_s\cdot\alpha$. Whether vortex shedding from the big jet pump opening occurs, is found to be dependent on the the local displacement amplitude at the position of the big opening $\xi_{1,b}$. A Keulegan--Carpenter number at the big jet pump opening is estimated using
\begin{equation}
KC_{D,b} = \frac{\xi_{1}\frac{A_s}{A_b}}{D_b} = \xi_{1}\frac{R_s^2}{2R_b^3}.
\label{eq:KCdb}
\end{equation}
Note that $\xi_{1}$ is still the local displacement amplitude at the jet pump waist. Fig.~\ref{fig:flowregimes_KCdb_x1Dsalpha} shows the flow regimes where either one-sided or two-sided vortex propagation occurs with $KC_{D,b}$ instead of $KC_{D}$ on the vertical axis. Only the range $0<\xi_1/D_s\cdot\alpha<0.7$ is shown, omitting the flow separation regime. The two-sided vortex propagation is first observed when $KC_{D,b}>0.15$. This is at a different value than where vortex propagation at the jet pump waist was first recognized ($KC_D>0.5$), which is an effect of the local dimensionless curvature $\chi$. The observed transition at $KC_{D,b}=0.15$ very nicely corresponds to the value found experimentally for a sharp-edged axisymmetric orifice.\cite{Holman2005} The effect of the jet pump curvature will be further investigated in Section~\ref{sec:var_Rc}.

\subsection{Jet pump performance}
\label{sec:varAlpha_varRs}

Ideally, a jet pump should impose the desired time-averaged pressure drop while having minimum acoustic power dissipation. Thus, the jet pump performance can be determined using two quantities: the time-averaged pressure drop $\Delta p_2$ and the acoustic power dissipation $\Delta\dot{E}_2$, both a function of the jet pump waist velocity amplitude $|u_{1,\mathit{JP}}|$. The time-averaged pressure drop over the jet pump is calculated from the simulation results by a subtraction of the mean pressure fields on either side of the jet pump. In this way the influence of the vortex propagation distance on the time-averaged pressure drop is eliminated.\cite{Oosterhuis2015}

The acoustic power is calculated using first order quantities as\cite{Swift2002_acousticpower}
\begin{equation}
\dot{E}_2(x) = \frac{1}{2}\Re{\left[\tilde{p}_1(x)U_1(x)\right]},
\label{eq:E2}
\end{equation}
with $\tilde{p}_1(x)$ the complex conjugate of the pressure amplitude area-averaged over the local cross section and $U_1(x)$ the complex volume flow rate. The dissipation of acoustic power across the jet pump, $\Delta\dot{E}_2$, is determined in a similar manner as the time-averaged pressure drop.

To better understand the relation between these two quantities, a scaling is applied according to the work of Smith and Swift:\cite{Smith2003a}
\begin{gather}
\Delta p_2^* = \frac{8\Delta p_2}{\rho_0 |u_{1,\mathit{JP}}|^2} ,
\label{eq:dp2_star}\\
\Delta \dot{E}_2^* = \frac{3\pi\Delta \dot{E}_2}{\rho_0\pi R_s^2 |u_{1,\mathit{JP}}|^3}.
\label{eq:dE2_star}
\end{gather}
Note that under the quasi--steady approximation, $\Delta p_2^*$ would represent the difference in minor loss coefficients between the two flow directions while $\Delta\dot{E}_2^*$ would represent the summation of the minor loss coefficients (i.e. the last terms in Eq.~\ref{eq:backhaus} and~\ref{eq:backhaus_dE}, respectively). Using these two dimensionless quantities, a jet pump effectiveness can be defined:\cite{Smith2003a}
\begin{equation}
\eta = \frac{\Delta p_2^*}{\Delta\dot{E}_2^*}.
\label{eq:eta}
\end{equation}
Optimal jet pump performance is achieved when $\eta$ is maximized. These three performance characteristics are calculated for each simulation case. In the following, they will be shown as a function of the scaling parameters introduced previously to study the effect of geometry changes on the jet pump performance. 

\subsubsection{Time-averaged pressure drop}
\begin{figure}
\centering
\subfloat[Dimensionless pressure drop as a function of $\xi_1/D_s\cdot\alpha$\label{fig:dp2_x1DsAlpha}]{\includegraphics[width=.5\textwidth]{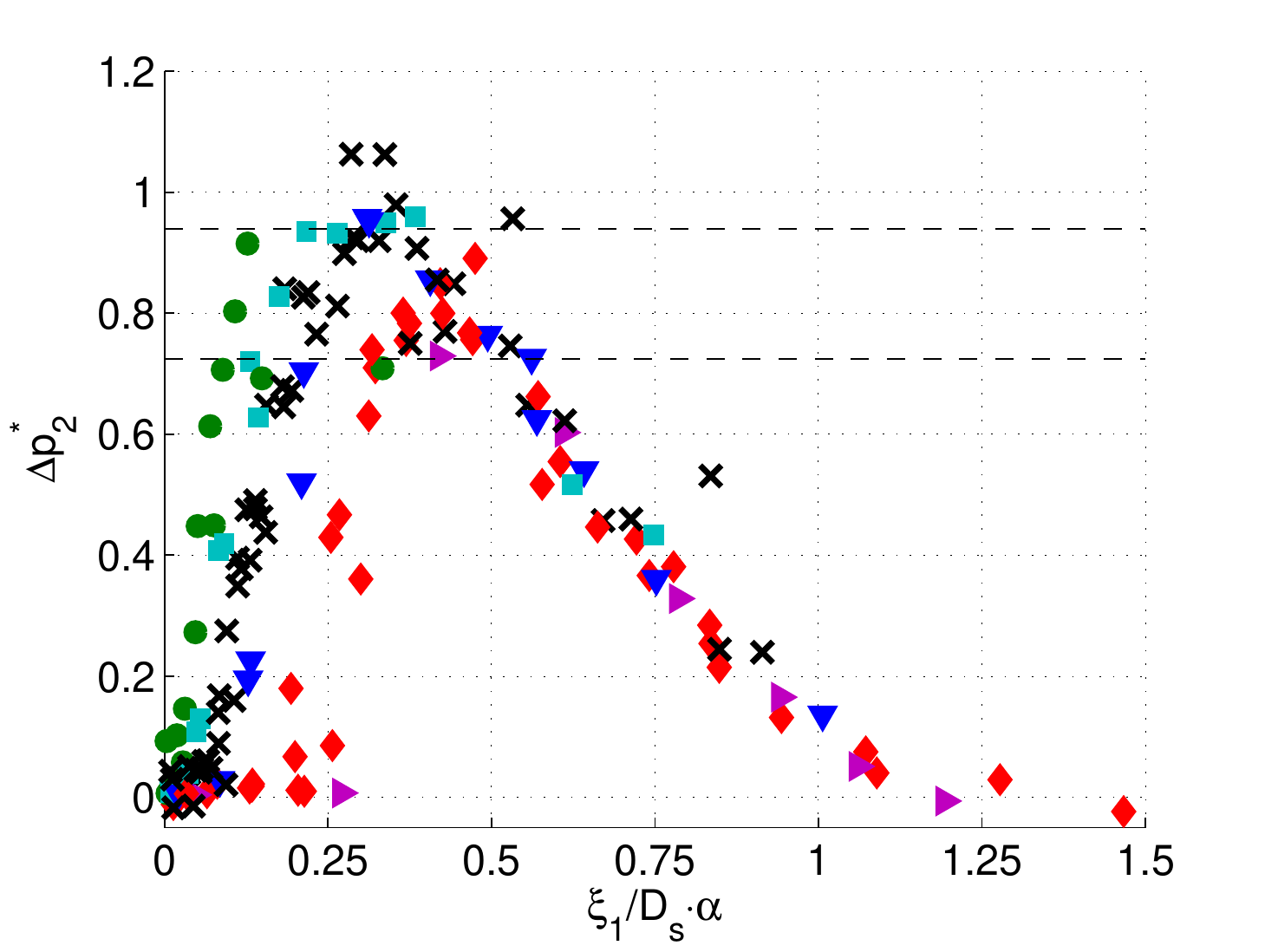}}
\subfloat[Dimensionless pressure drop as a function of $KC_D$, only the range $KC_D<3$ is shown.\label{fig:dp2_KCd}]{\includegraphics[width=.5\textwidth]{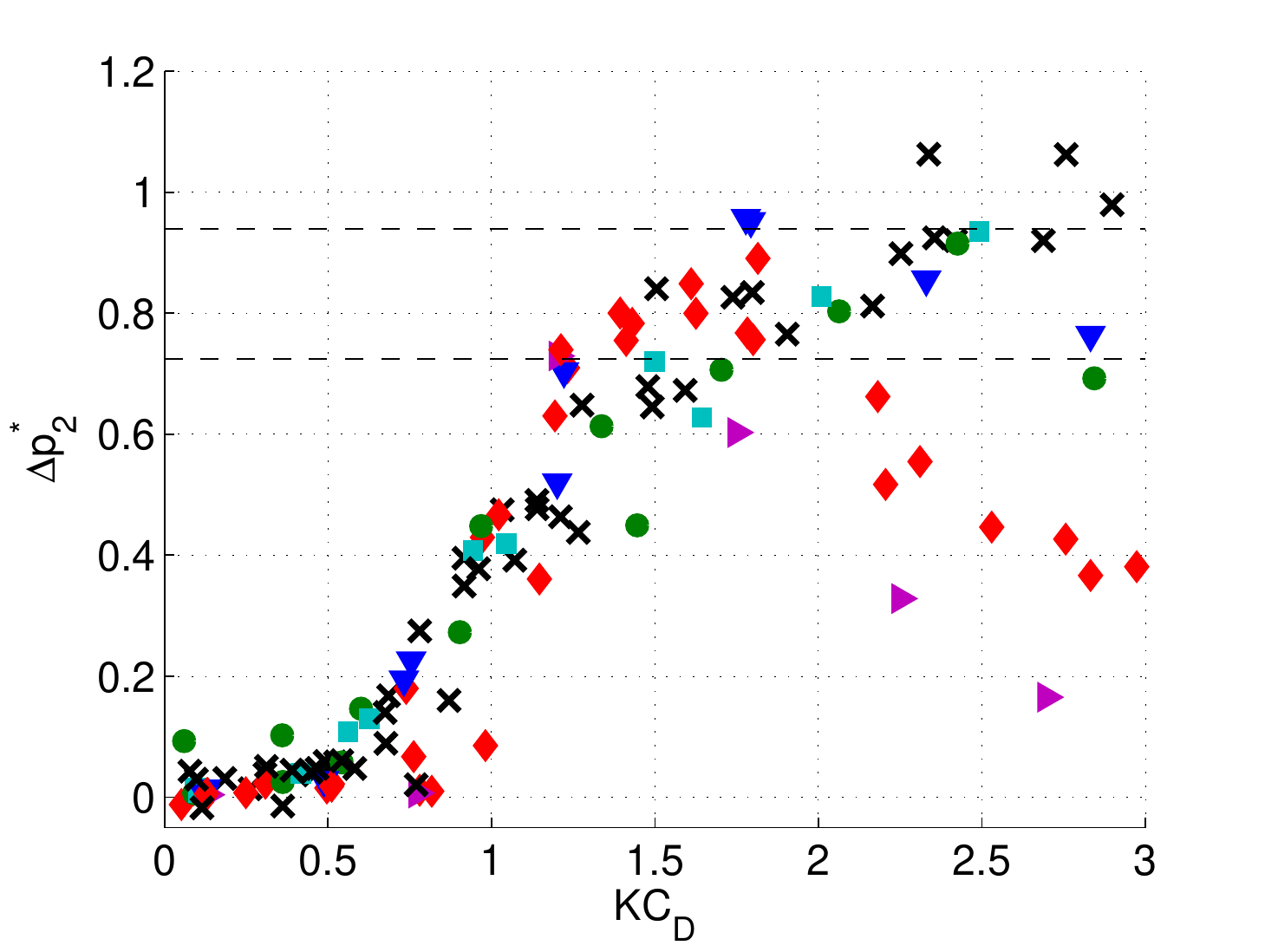}}
\caption{Dimensionless pressure drop using jet pump geometries with varied taper angle (Table~\ref{tab:geom_varalpha}) or waist diameter (Table~\ref{tab:geom_varRs}). The symbols represent different taper angles as listed in Table~\ref{tab:geom_varalpha}. The thin dashed lines indicate the upper and lower values of $\Delta p_2^*$ from the quasi--steady approximation (Eq.~\ref{eq:backhaus}) for the various geometries (color online).}
\label{fig:dp2_varAlpha+varRs}
\end{figure}

Fig.~\ref{fig:dp2_x1DsAlpha} shows the dimensionless pressure drop $\Delta p_2^*$ as a function $\xi_1/D_s\cdot\alpha$. The symbols represent different taper angles, obtained by either changing the jet pump length or the big radius, each depicted in Table~\ref{tab:geom_varalpha}. The results with a varied waist diameter are all plotted using crosses ($\times$), representing their taper angle of $\alpha=\SI{7}{\degree}$. The maximum dimensionless pressure drop for all cases is achieved between $0.2<\xi_1/D_s\cdot\alpha<0.5$ which fits well within the determined range of the flow regimes where vortex shedding, but no flow separation, occurs. This emphasizes the direct relation between the flow field asymmetry and the time-averaged pressure drop. 

A decay in $\Delta p_2^*$ is observed for higher values of $\xi_1/D_s\cdot\alpha$. The range where flow separation was observed coincides with this decay in dimensionless pressure drop. The decay follows the same trend for all jet pump geometries and is properly scaled using the adjusted Keulegan--Carpenter number, which shows the dependency of the flow separation occurrence on the taper angle. Moreover, a corresponding decay in $\Delta p_2^*$ as a function of $\xi_1/D_s\cdot\alpha$ is observed in preliminary experimental results that have been obtained at $Re<Re_c$ using the \SI{15}{\degree} jet pump geometry in an oscillatory flow facility.\cite{Vidya2014,Aben2010} By performing hot-wire anemometry near the big jet pump opening the existence of an outward directed vortex street was revealed when $\xi_1/D_s\cdot\alpha>0.7$, confirming the observations from the current simulations. A detailed discussion of the experimental results is deferred to a future publication.

In the published data of Petculescu and Wilen however, the decay in dimensionless pressure drop is less severe.\cite{Petculescu2003} Their values of $\Delta p_2^*$ typically stabilize between 0.20 and 0.50 at high values of $\xi_1/D_s\cdot\alpha$, depending on the taper angle used, and do not decrease to zero. Two possible explanations will be discussed: their small waist diameter compared to the viscous penetration depth and a transition to turbulence. In the geometries of Petculescu and Wilen, $D_s/\delta_\nu \approx 8.3$ while for the current geometries $D_s/\delta_\nu$ ranges from \SIrange{27}{98}{}. When the waist diameter is on the order of the viscous penetration depth, the occurrence of flow separation, and consequently the dimensionless pressure drop at $\xi_1/D_s\cdot\alpha>0.7$,  is most likely affected due to viscous effects dominating the flow and preventing the flow from separating. Alternatively, turbulence can affect the decay in $\Delta p_2^*$ when the turbulent boundary layer prevents the flow from separating.\cite{Schlichting1968_separation} A detailed investigation of the effect of turbulence on the oscillatory flow in jet pumps is subject to further research. 

The adjusted Keulegan--Carpenter number, $\xi_1/D_s\cdot\alpha$, provides consistent scaling of the dimensionless pressure drop in the decaying part of the curve shown in Fig.~\ref{fig:dp2_x1DsAlpha}. However, a large spread from the reference geometry is observed for low values of the adjusted Keulegan--Carpenter number ($\xi_1/D_s\cdot\alpha<0.2$). Fig.~\ref{fig:dp2_KCd} shows the dimensionless pressure drop as a function of $KC_D$ which better scales the increase of $\Delta p_2^*$ towards its maximum. This shows that the initial increase of $\Delta p_2^*$ towards its maximum is not dependent on the taper angle at all. A clear similarity in this graph is the range $KC_D<0.5$ where the pressure drop is negligible, corresponding to the flow regime with oscillatory vortices shown in Fig.~\ref{fig:flowregimes_KCd_x1DsAlpha}. At larger values of $KC_D$, the effect of the taper angle becomes apparent and the results of the different geometries deviate from each other.

\subsubsection{Acoustic power dissipation}
\begin{figure}
\centering
\includegraphics[width=.5\textwidth]{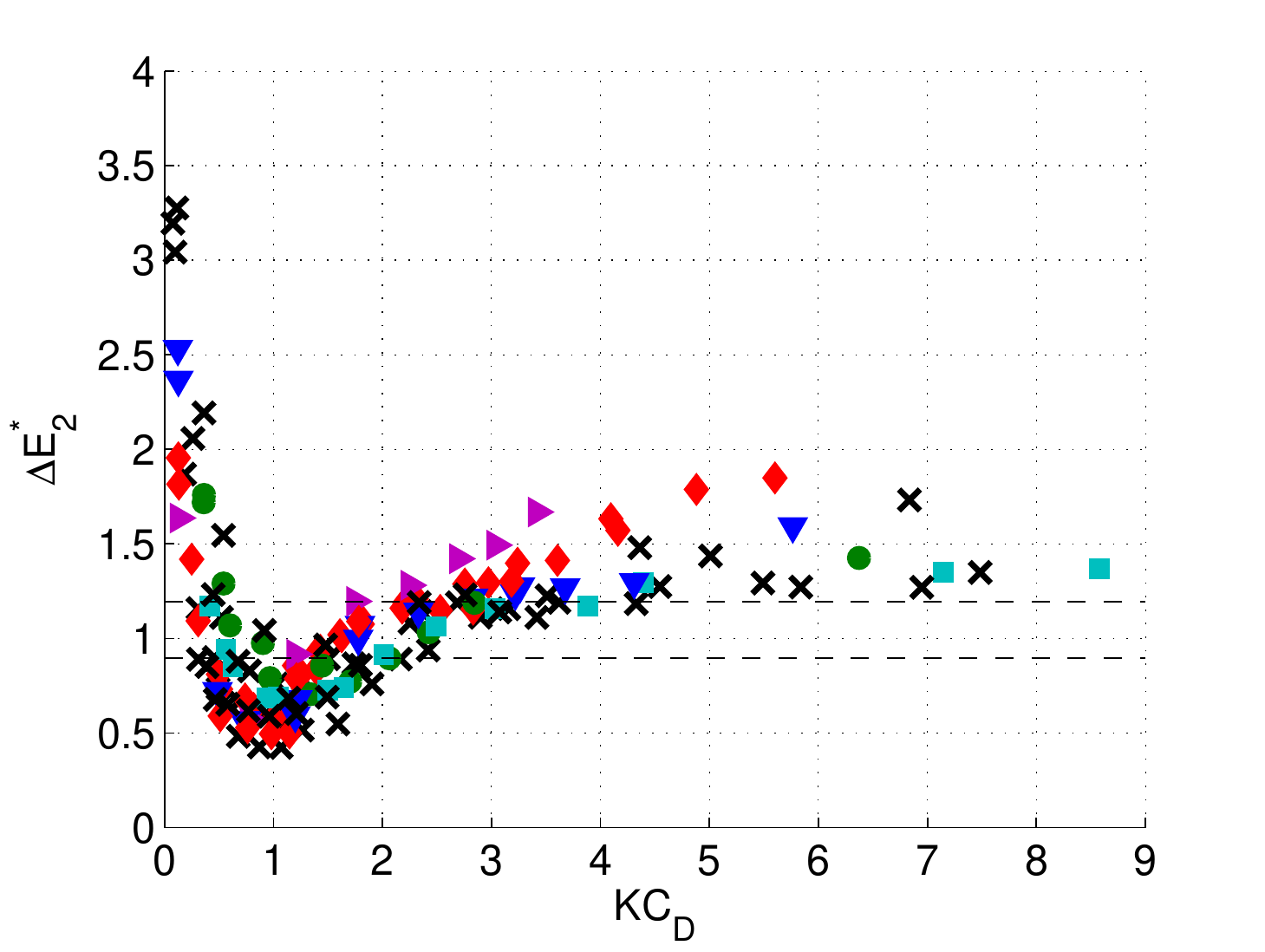}
\caption{Dimensionless acoustic power dissipation using jet pump geometries with varied taper angle (Table~\ref{tab:geom_varalpha}) or waist diameter (Table~\ref{tab:geom_varRs}). Symbols in accordance with Fig.~\ref{fig:dp2_varAlpha+varRs} (color online). The thin dashed lines indicate upper and lower values of the predicted acoustic power dissipation according to the quasi--steady approximation (Eq.~\ref{eq:backhaus_dE}).}
\label{fig:dE2_KCd}
\end{figure}

By studying the dimensionless acoustic power dissipation, it is observed that the jet pump taper angle has less effect on this performance quantity compared to the pressure drop. Instead, the dimensionless acoustic power dissipation ($\Delta\dot{E}_2^*$) shows good similarity between the different jet pump geometries when scaled using $KC_D$. Fig.~\ref{fig:dE2_KCd} shows the dimensionless acoustic power dissipation as a function of the Keulegan--Carpenter number based on the jet pump waist diameter, $KC_D$. The symbols correspond to those in Fig.~\ref{fig:dp2_varAlpha+varRs}. A common minimum occurs around $KC_D \approx 1$ while at higher amplitudes ($KC_D>3$) $\Delta\dot{E}_2^*$ only slightly increases. Although the taper angle has no effect on the location of the minimum dimensionless acoustic power dissipation, the slight increase of $\Delta\dot{E}_2^*$ at higher amplitudes is easily observed for higher taper angles (\SI{15}{\degree}, $\blacklozenge$ and \SI{20}{\degree}, $\blacktriangleright$).

\subsubsection{Comparison to quasi--steady model}
The obtained simulation results can all be compared to the quasi--steady model proposed by Backhaus and Swift.\cite{Backhaus2000} In this model, the values of $\Delta p_2^*$ and $\Delta\dot{E}_2^*$ are assumed to be amplitude independent and the only significant variation in the quasi--steady model occurs by changing the waist diameter. In Fig.~\ref{fig:dp2_varAlpha+varRs}--\ref{fig:dE2_KCd} the upper and lower bounds of these quasi--steady approximated values are indicated by the thin dashed lines. It becomes clear that the quasi--steady approximation is a good approach to estimate the time-averaged pressure drop of a jet pump when operating in the optimum ranges of $\xi_1/D_s\cdot\alpha$ and $KC_D$. Outside of these ranges however, the actual performance will deviate significantly from the predicted values. The simulated acoustic power dissipation is especially in good agreement for the lower taper angles and $KC_D>2$. For high taper angles (\SI{15}{\degree}, $\blacklozenge$ and \SI{20}{\degree}, $\blacktriangleright$) however, a deviation from the quasi--steady model is observed at $KC_D>2$. On average the quasi--steady values show considerable correspondence to the measured dimensionless acoustic power dissipation. 

\subsubsection{Jet pump effectiveness}
\begin{figure}
\centering
\includegraphics[width=.5\textwidth]{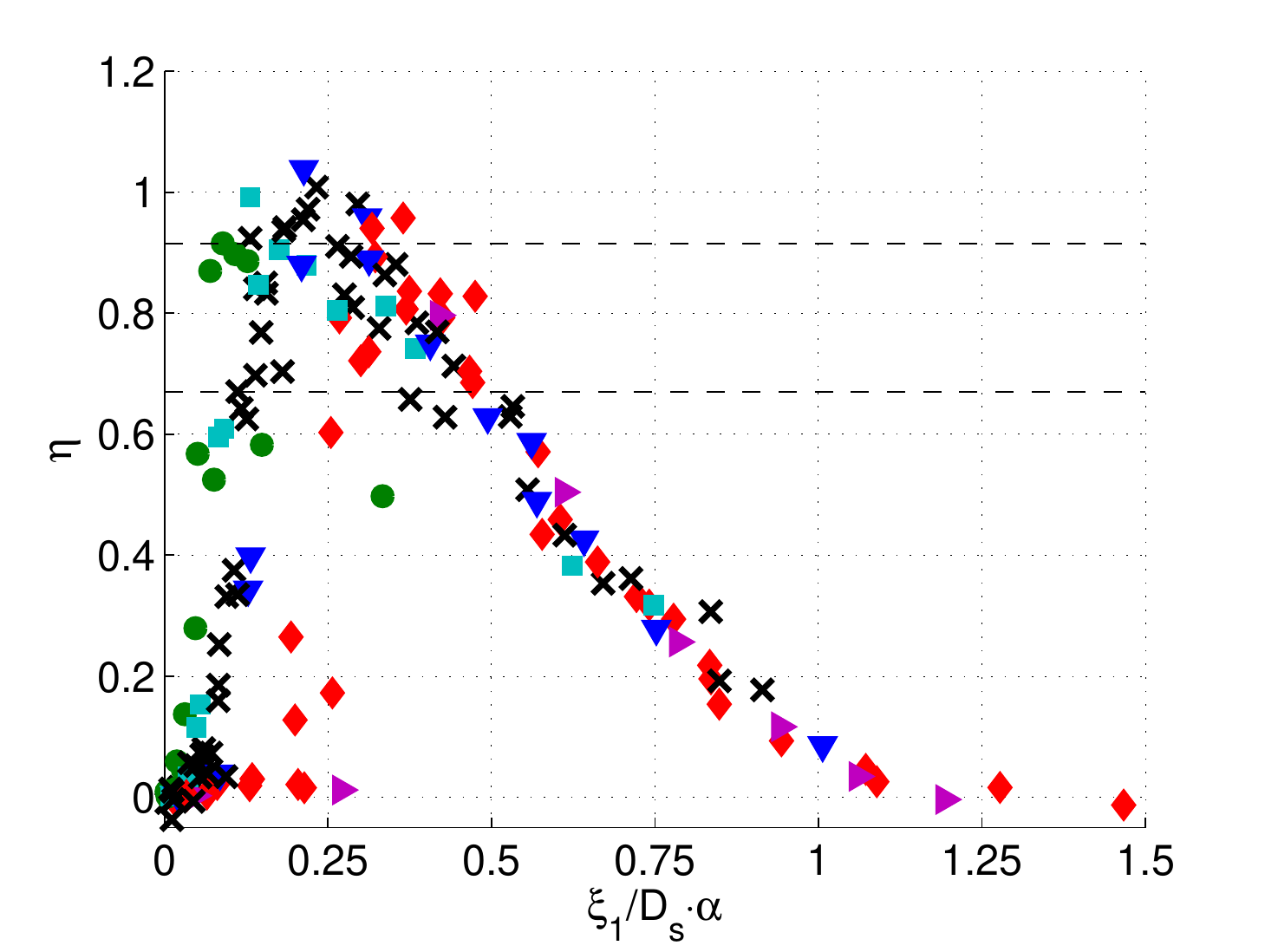}
\caption{Jet pump effectiveness for geometries with varied taper angle (Table~\ref{tab:geom_varalpha}) or waist diameter (Table~\ref{tab:geom_varRs}). Symbols in accordance with Fig.~\ref{fig:dp2_varAlpha+varRs} (color online).}
\label{fig:eta_x1DsAlpha}
\end{figure}

The two jet pump performance characteristics, dimensionless pressure drop and acoustic power dissipation, can be combined into one effectiveness parameter $\eta$ as defined in Eq.~\ref{eq:eta}. Fig.~\ref{fig:eta_x1DsAlpha} shows the jet pump effectiveness for all geometries with varied taper angle or waist diameter. The highest jet pump effectiveness is again obtained between $0.2<\xi_1/D_s\cdot\alpha<0.5$. Note that this does not strictly coincide with the minimum in the acoustic power dissipation (at $KC_D\approx 1$). However, as long as $KC_D>0.5$, the variations in dimensionless pressure drop have a larger influence on the jet pump effectiveness than the relative small changes in the dimensionless acoustic power dissipation.
\\

Summarizing the results so far, it can be concluded that for laminar oscillatory air flows the optimum jet pump performance is achieved between $0.2<\xi_1/D_s\cdot\alpha<0.5$ under the condition that $KC_D>0.5$. The combination of these two criteria imposes a limit on the maximum taper angle which maximizes the dimensionless pressure drop.  Substitution of the two criteria, yields $\alpha<\SI{23}{\degree}$ in order to prevent overlapping of the range of optimum jet pump performance with the range where no pressure drop is measured. The optimum jet pump performance criterion does well coincide with the region where either right-sided or two-sided vortex shedding is observed (indicated with $\blacksquare$ and $\blacktriangleright$ in Fig.~\ref{fig:flowregimes_KCd_x1DsAlpha}). It is only in this optimum region that the measured dimensionless pressure drop will approach values as predicted by the quasi--steady model.

\subsection{Influence of waist curvature}
\label{sec:var_Rc}

In the previous discussion where the influence of the jet pump taper angle and waist diameter were investigated, the radius of curvature at the jet pump waist was kept constant. In steady flow theory, the minor loss coefficient for contraction is dependent on the dimensionless radius of curvature,\cite{Idelchik2007} $\chi=R_c/D_s$. In order to maximize the time-averaged pressure drop while minimizing the acoustic power dissipation, it is desirable to minimize the contraction minor loss coefficient (following Eq.~\ref{eq:backhaus}--\ref{eq:backhaus_dE}) and as such, apply a smooth curvature at the jet pump waist. For a ``smooth'' contraction ($\chi>0.15$) the theoretical minor loss coefficient is $K_\mathit{con}=0.04$ which increases up to $K_\mathit{con}=0.50$ for a sharp contraction.\cite{Idelchik2007} For all the jet pump geometries shown so far, a curvature of $\chi=0.36$ was applied. In the current section, the jet pump curvature will be varied to study both its effect on the jet pump performance as well as on the occurring flow phenomena.

Six different dimensionless curvatures have been investigated, ranging from $\chi=$~\SIrange{0}{0.65}{}. Aside from the radius of curvature $R_c$, the jet pump geometry is equal to the reference geometry found in Table~\ref{tab:geom_ref}. An overview of the studied geometries with variable curvature is shown in Table~\ref{tab:geom_varRc}.
\begin{table}
\centering
	\caption{Dimensions of jet pump geometries with varied radius of curvature $R_c$. Constant taper angle $\alpha=\SI{7}{\degree}$. Horizontal line separates cases at $\chi=0.15$.}
	\label{tab:geom_varRc}
	\begin{tabular}{llll}
	No. & $R_c$ & $\chi$ &\\
	\hline
	14 & \SI{0.0}{\mm} & 0 & \multirow{2}{*}{``sharp''}\\
	15 & \SI{1.0}{\mm} & 0.08 & \\
	\hline
	16 & \SI{2.5}{\mm} & 0.19 & \multirow{4}{*}{``smooth''}\\
	\textit{ref} & \SI{5.0}{\mm} & 0.36 & \\
	17 & \SI{7.0}{\mm} & 0.48 & \\
	18 & \SI{10.0}{\mm} & 0.65 & \\

	\end{tabular}
\end{table}

\subsubsection{Jet pump performance}
\begin{figure}
\centering
\subfloat[\label{fig:dp2_KCd[varRc]}]{\includegraphics[width=.5\textwidth]{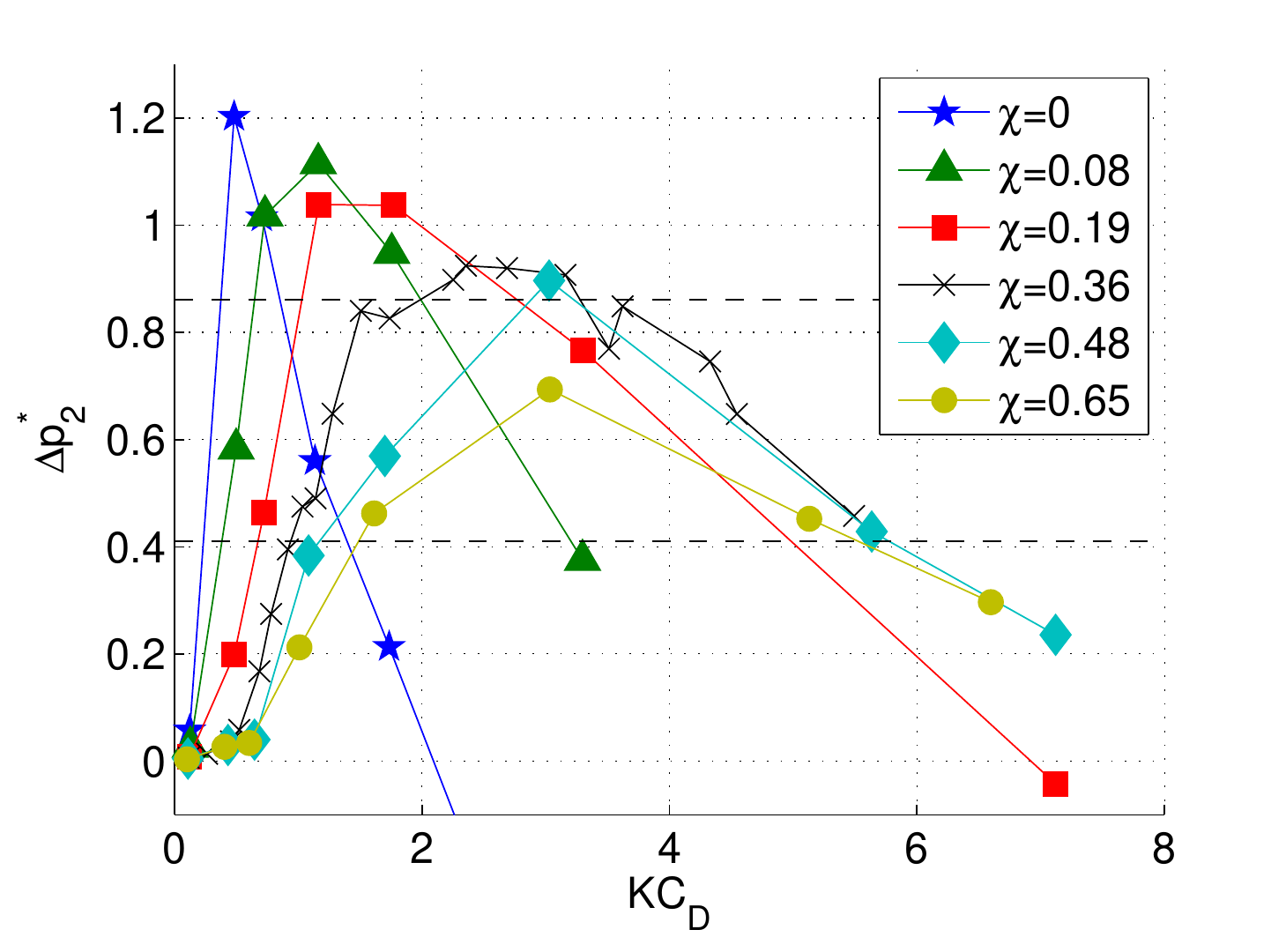}}
\subfloat[\label{fig:dp2_KCd-alt1[varRc]}]{\includegraphics[width=.5\textwidth]{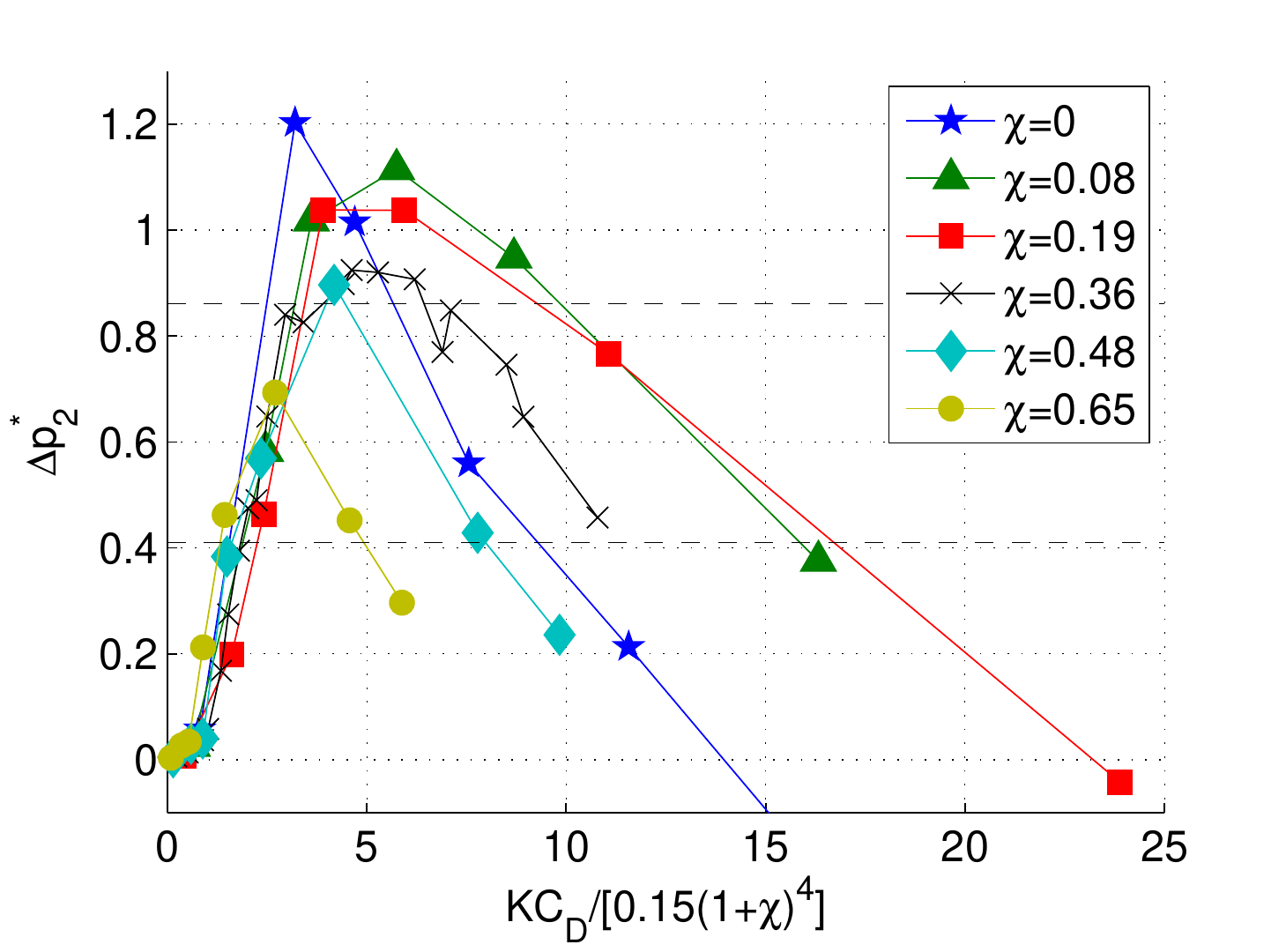}}\\
\subfloat[\label{fig:dp2_KCd-alt2[varRc]}]{\includegraphics[width=.5\textwidth]{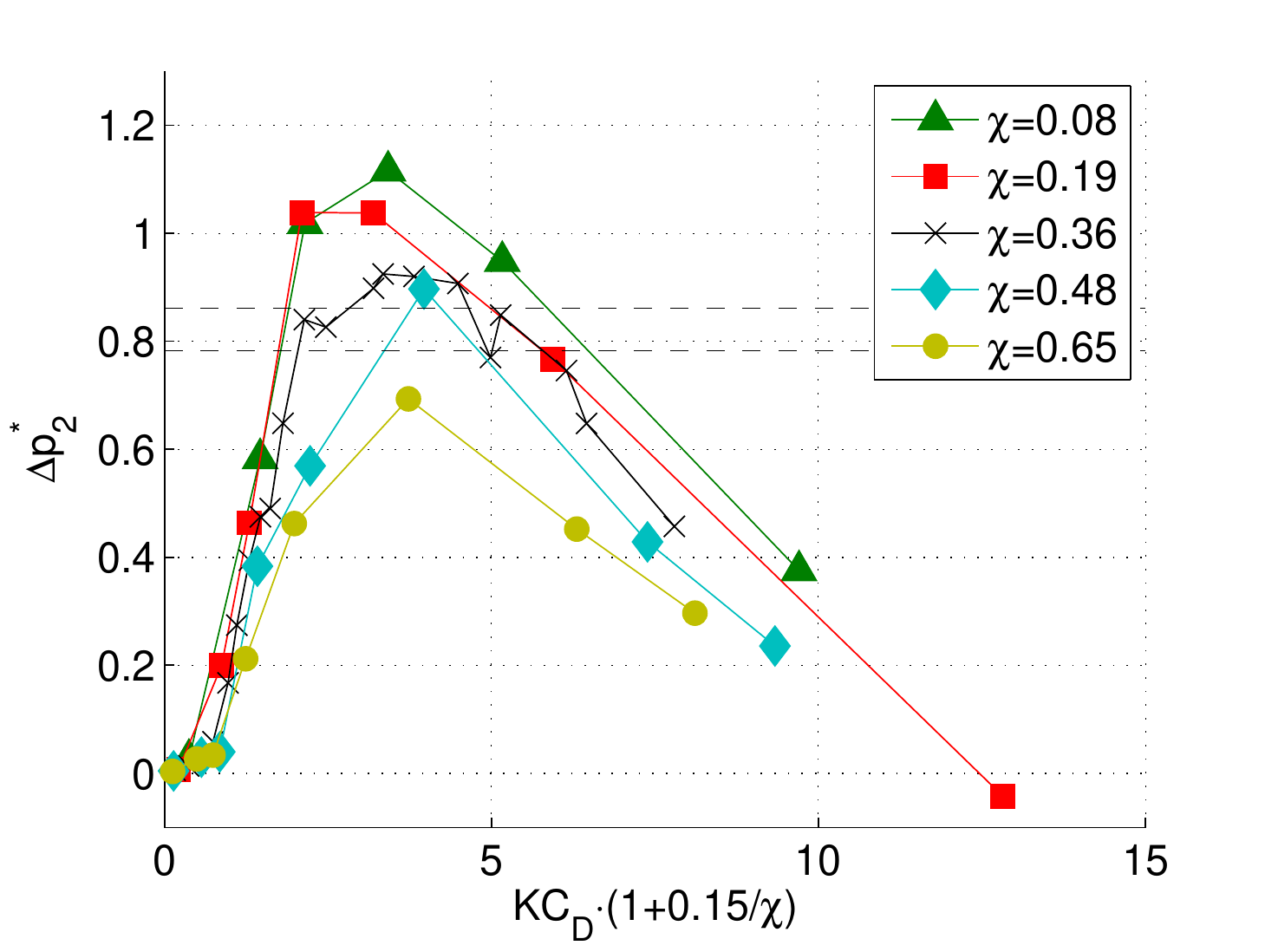}}
\caption{Dimensionless pressure drop using jet pump geometries with varied waist curvature as a function of three different scaling parameters: a) $KC_D$, b) $KC_D /0.15(1+\chi)^4$ and c) $KC_D\cdot(1+0.15/\chi)$. The different symbols correspond to dimensionless curvatures ranging from $\chi=$~\SIrange{0}{0.65}{}. The dashed black lines indicate the upper and lower values from the quasi--steady approximation for the various values of $\chi$ (color online).}
\label{fig:varRc_dp2}
\end{figure} 
When comparing the measured dimensionless pressure drop using these jet pump samples, it is found that an additional contribution to the Keulegan--Carpenter number is required to properly scale the results from different curvatures. Fig.~\ref{fig:dp2_KCd[varRc]} shows the dimensionless pressure drop as a function of $KC_D$ for the geometries with varied curvature.  The investigated curvatures are indicated by different symbols depicted in the corresponding legend. The crosses ($\times$) represent the reference geometry where $\chi=0.36$. The shape of the dimensionless pressure drop curve is found to be inversely proportional to the dimensionless curvature, especially for low values of $\chi$. For large curvatures, the effect on the dimensionless pressure drop becomes less apparent. Holman et al. studied the effect of edge curvature on the onset of a synthetic jet.\cite{Holman2005} A formation criterion of the form $KC_D>C\cdot(1+\chi)^p$ was derived based on the vorticity flux through the jet. The applicability of this criterion to jet pumps with variable curvature is investigated in the following. It is observed that especially the initial increase in $\Delta p_2^*$, corresponding to the onset of vortex shedding, can be properly scaled using $KC_D / C (1+\chi)^p$ with $p=4$. Given that the onset of vortex shedding for a jet pump geometry with $\chi=0.36$ is found at $KC_D=0.5$, the additional scaling parameter $C = 0.15$ can be estimated. Fig.~\ref{fig:dp2_KCd-alt1[varRc]} shows the dimensionless pressure drop as a function of the adjusted scaling parameter. While the initial increase in $\Delta p_2^*$ for the various curvatures all collapse to a single curve, the decay in $\Delta p_2^*$ due to the occurrence of flow separation cannot properly be accounted for by any scaling of the form $KC_D/C(1+\chi)^p$.

To overcome this, an alternative scaling strategy for $\chi$ is proposed of the form $KC_D\cdot(1+a/\chi)$. Using a least--squares approach, the fitting parameter $a=0.15$ is determined. Note that $0.15/\chi$ represents the fraction of the dimensionless curvature with respect to a ``smooth'' contraction. Fig.~\ref{fig:dp2_KCd-alt2[varRc]} shows again the dimensionless pressure drop data, but now plotted against $KC_D\cdot(1+0.15/\chi)$. The case with no curvature ($\chi=0$) does not appear in this figure as it will lead to infinite values on the horizontal axis. Both the maxima of the dimensionless pressure drop as well as the decay at higher amplitudes are well aligned using this alternative scaling approach.

What is remarkable is that the maximum measured dimensionless pressure drop decreases with increasing curvature, which is in contrast with the steady flow theory where the minor loss coefficient for contraction is expected to fall until $\chi>0.15$. The maximum achieved dimensionless pressure drop ranges from $\Delta p_2^*=1.20$ for a sharp contraction ($\chi=0$) to $\Delta p_2^*=0.69$ for $\chi=0.65$ in an almost linear manner. Furthermore, the effect of curvature on the dimensionless pressure drop is not only reversed, but the variation is larger than what is predicted by the quasi--steady model (dashed lines). An explanation of this behavior is the effect of curvature on the expansion phase, which is typically neglected in a quasi--steady approximation of the jet pump performance. Not only will the minor loss coefficient for contraction at the jet pump waist decrease, but so does the minor loss coefficient for expansion. This was first experimentally determined by Smith under steady flow conditions.\cite{Smith2004} A decrease in the expansion minor loss coefficient at the jet pump waist will lead to a decrease in the dimensionless pressure drop according to Eq.~\ref{eq:backhaus}. 

\begin{figure}
\centering
\subfloat[\label{fig:dE2[varRc]}]{\includegraphics[width=.5\textwidth]{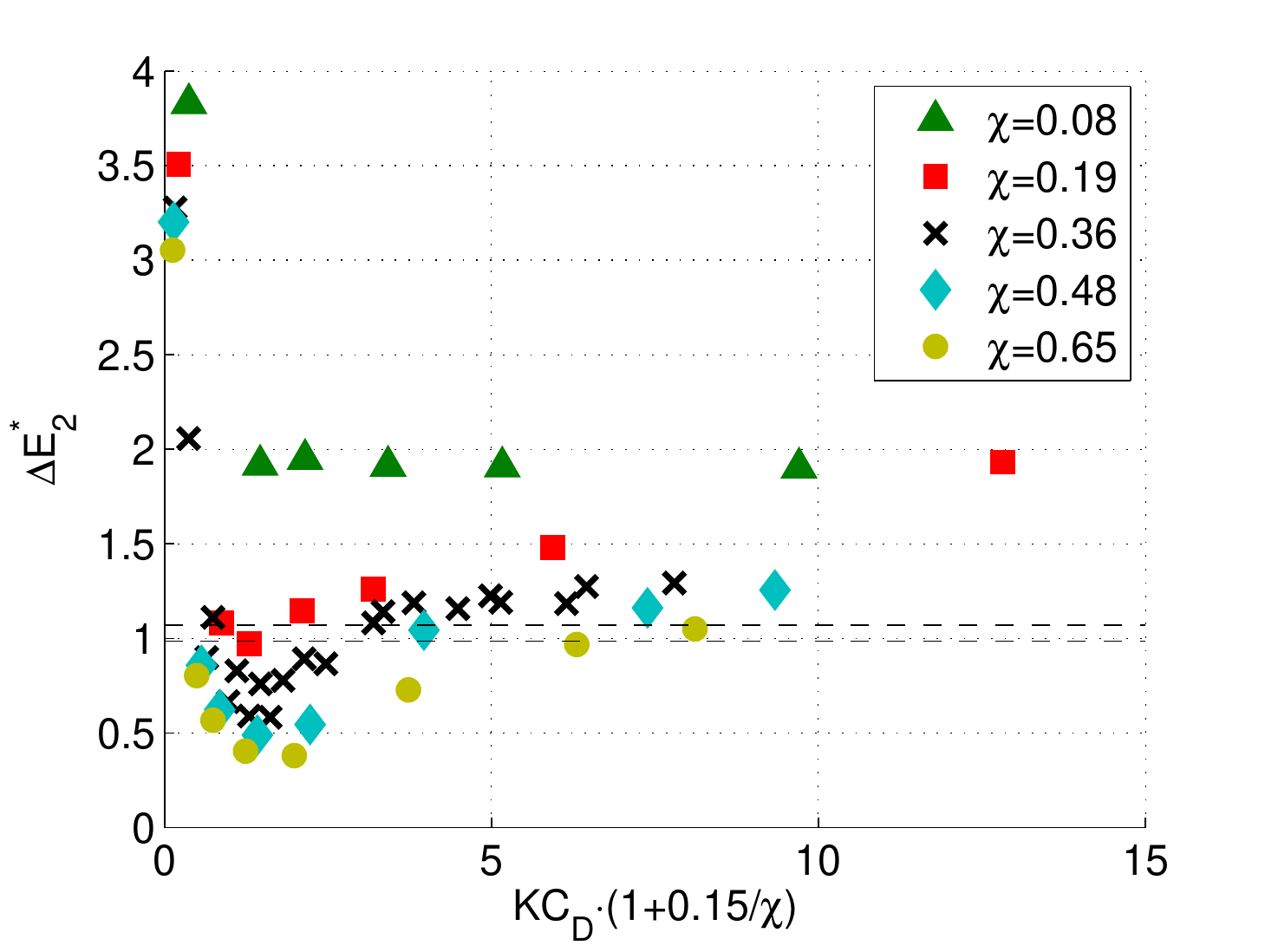}}
\subfloat[\label{fig:eta[varRc]}]{\includegraphics[width=.5\textwidth]{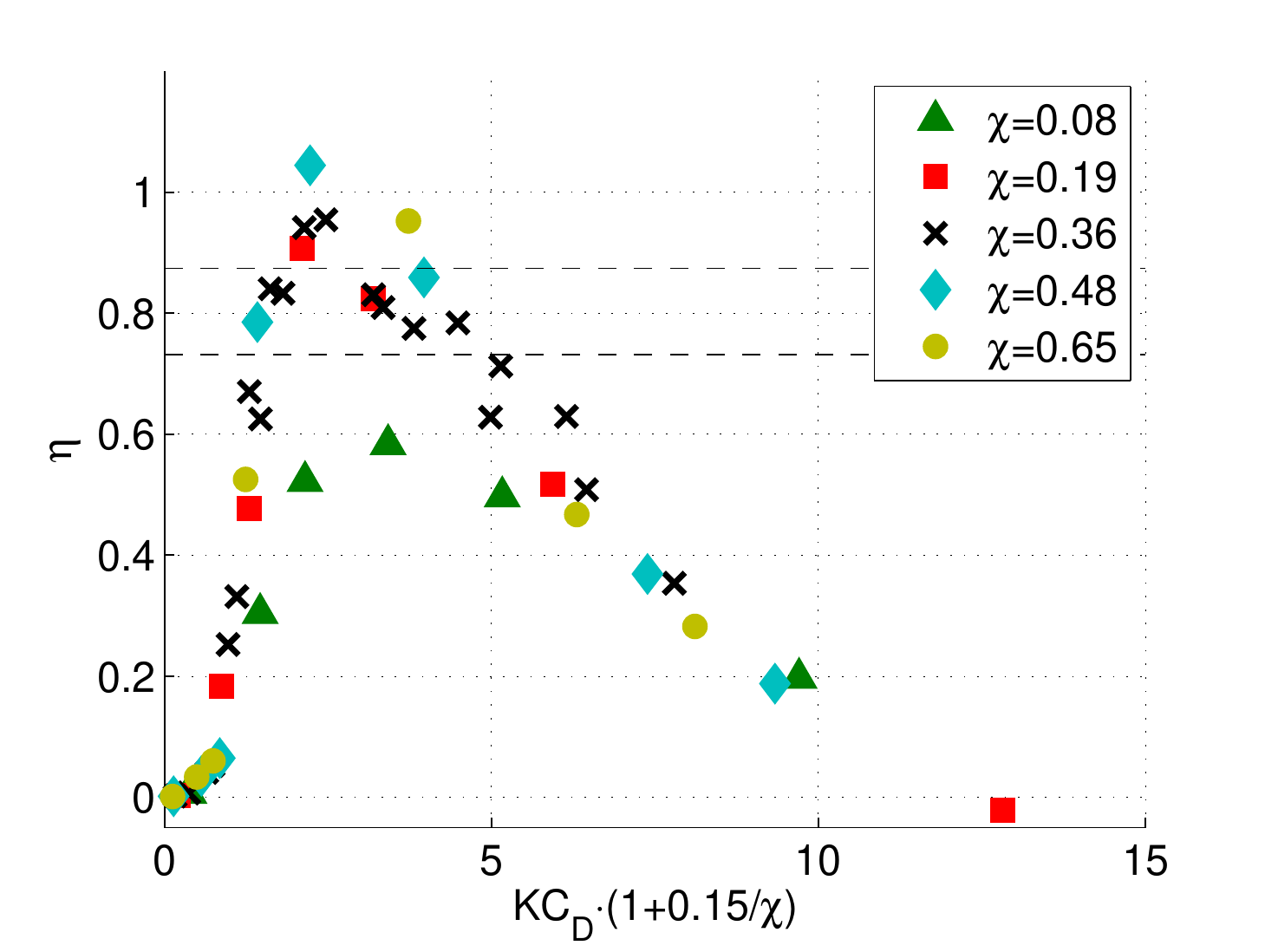}}
\caption{Dimensionless acoustic power dissipation (a) and jet pump effectiveness (b) using jet pump geometries with varied waist curvature and scaled using $KC_D\cdot(1+0.15/\chi)$. The different symbols correspond to dimensionless curvatures ranging from $\chi=$~\SIrange{0.08}{0.65}{}. The dashed black lines indicate the upper and lower values from the quasi--steady approximation for the various values of $\chi$ (color online).}
\label{fig:varRc_performance}
\end{figure}

Additionally, this is reflected in the measured dimensionless acoustic power dissipation, which is shown in Fig.~\ref{fig:dE2[varRc]}. The minimum achieved dimensionless acoustic power dissipation decreases more than six times between $\chi=0$ (not shown) and $\chi=0.65$. A similar effect has been reported when rounding the orifice of an synthetic jet.\cite{Nani2012} Combining the effect the curvature has on both the pressure drop and acoustic power dissipation using the jet pump effectiveness, shows that a higher dimensionless curvature leads to a better jet pump performance. This is depicted in Fig.~\ref{fig:eta[varRc]} where the jet pump effectiveness for five investigated curvatures is shown.

\subsubsection{Flow regimes}
\begin{figure}
\centering
\includegraphics[width=.5\textwidth]{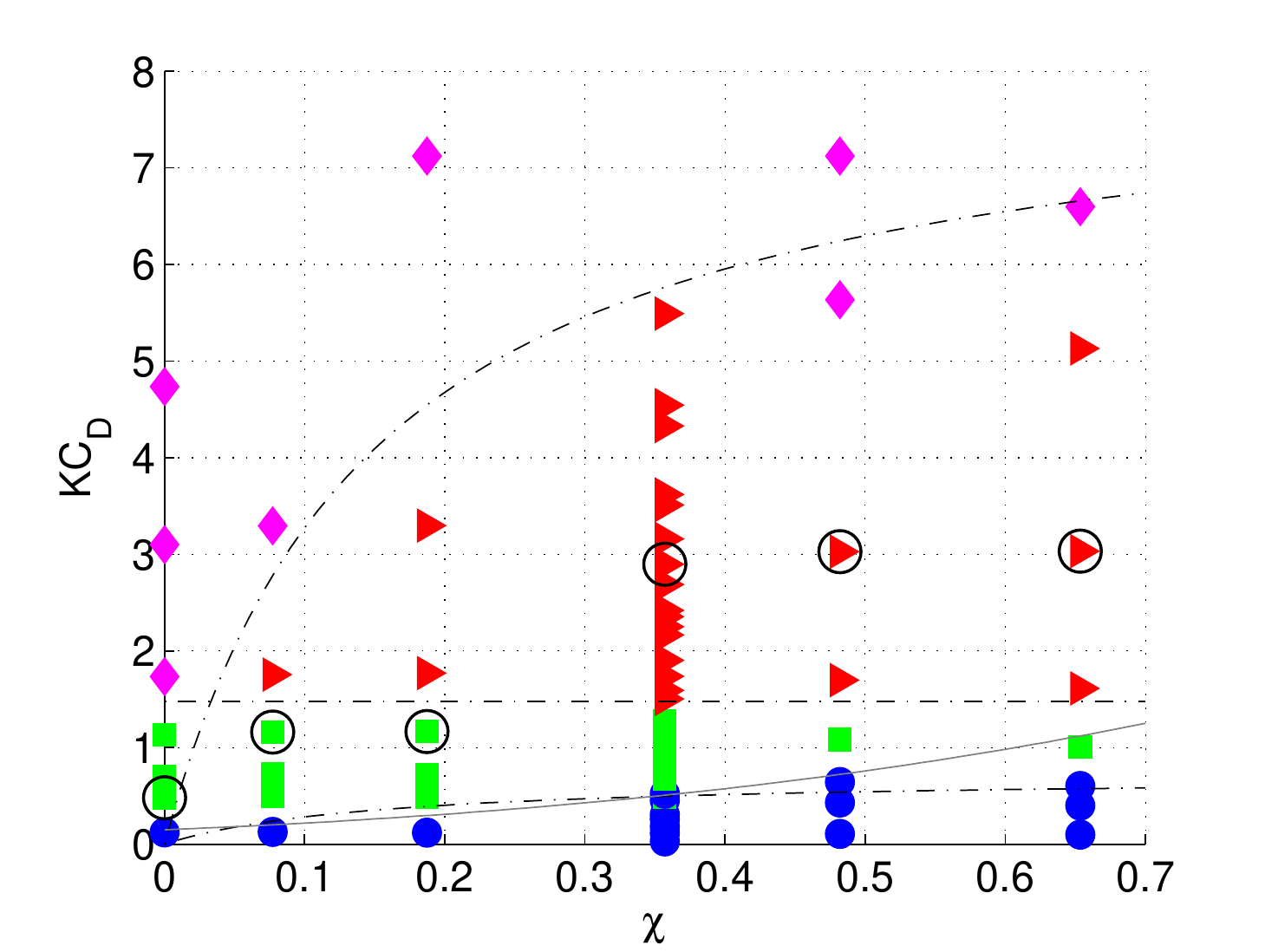}
\caption{Distribution of the different flow regimes in the ($\chi$, $KC_D$) space for jet pump geometries with varied curvature (Table~\ref{tab:geom_varRc}) and a \SI{7}{\degree} taper angle. Symbols are in accordance with Fig.~\ref{fig:flowregimes_varAlpha+varRs} with the open circles showing the position where $\Delta p_2^*$ is maximum for each investigated curvature. The dashed lines represent the bounds of the different flow regimes determined according to Eq.~\ref{eq:flowbound1}--\ref{eq:flowbound4} and the gray solid line represents the onset of vortex shedding according to Eq.~\ref{eq:flowbound1a} (color online).}
\label{fig:varRc_flowregimes}
\end{figure}

Similar to what was presented in Section~\ref{sec:flowregimes}, the observed jet pump behavior can be explained by studying the flow field and distinguishing between the four different flow regimes. Fig.~\ref{fig:varRc_flowregimes} shows the distribution of the flow regimes in the ($\chi$, $KC_D$) space. The four flow regimes are represented by different symbols in accordance to the symbols used in Fig.~\ref{fig:flowregimes_varAlpha+varRs}. A clear influence of the dimensionless curvature is visible on the value of $KC_D$ where vortex shedding is first observed (transition from {\Large$\bullet$} to $\blacksquare$) as well as on the value where flow separation is initiated (transition from $\blacktriangleright$ to $\blacklozenge$). The boundary between one-sided and two-sided vortex shedding (transition from $\blacksquare$ to $\blacktriangleright$) is not determined by the jet pump waist curvature at all. As the local curvature at the big opening is not changed, no dependency on $\chi$ is to be expected.

By substituting the original defined flow regime bounds from Section~\ref{sec:flowregimes} in the adjusted scaling parameter $KC_D\cdot(1+0.15/\chi)$, the bounds between the flow regimes (indicated by symbols as used in Fig.~\ref{fig:flowregimes_varAlpha+varRs} and Fig.~\ref{fig:varRc_flowregimes}) can be represented in the ($\chi$, $KC_D$) space:
\begin{eqnarray}
\text{{\Large$\bullet$} $\to$ $\blacksquare$: } KC_D &=& 0.7\left(1+\frac{0.15}{\chi}\right)^{-1} \label{eq:flowbound1}, \\
\text{$\blacksquare$ $\to$ $\blacktriangleright$: } KC_D &=& 0.15\frac{R_b^3}{R_s^3} \label{eq:flowbound2}, \\
\text{$\blacktriangleright$ $\to$ $\blacklozenge$: } KC_D &=& \frac{1}{\alpha}\left(1+\frac{0.15}{\chi}\right)^{-1} \label{eq:flowbound4},
\end{eqnarray}
where the first expression corresponds to $KC_D=0.5$, the second expression corresponds to $KC_{D,b}=0.15$ and the last expression corresponds to $\xi_1/D_s\cdot\alpha=0.7$ for the cases where $\chi=0.36$. The bounds are shown in Fig.~\ref{fig:varRc_flowregimes} by the dashed lines. The bottom curve represents the initiation of vortex shedding (Eq.~\ref{eq:flowbound1}) and the upper curve the transition to flow separation (Eq.~\ref{eq:flowbound4}), both a function of $\chi$. The horizontal dashed line in Fig.~\ref{fig:varRc_flowregimes} indicates the boundary between one-sided and two-sided vortex shedding. 

Based on the data available the determined bounds do separate the flow regimes depicted in Fig.~\ref{fig:varRc_flowregimes} well, although the defined onset of vortex shedding ({\Large$\bullet$} $\to$ $\blacksquare$) is subject to discussion. Given the fact that the initial increase in $\Delta p_2^*$ caused by the onset of vortex shedding was well described by adopting a scaling of the form proposed by Holman et al.,\cite{Holman2005} a flow regime bound based on the latter could be more appropriate:
\begin{equation}
\text{{\Large$\bullet$} $\to$ $\blacksquare$: } KC_D = 0.15\left(1+\chi\right)^4.
\label{eq:flowbound1a}
\end{equation}
This is represented in Fig.~\ref{fig:varRc_flowregimes} by a gray solid line and does indeed separate the two flow regimes. It should be noted that the data available is not sufficient to determine whether Eq.~\ref{eq:flowbound1} or Eq.~\ref{eq:flowbound1a} is more appropriate. 

Nevertheless, Eq.~\ref{eq:flowbound1a} predicts vortex shedding for $\chi=0$ at $KC_D>0.15$ which is also the formation criterion observed experimentally for a sharp edged synthetic jet.\cite{Holman2005} This exactly describes the difference in the transitional Keulegan--Carpenter number, observed in Section~\ref{sec:flowregimes}, between vortex shedding from the jet pump waist ($KC_{D,s}>0.5$) and vortex shedding from the big jet pump opening ($KC_{D,b}>0.15$). A lower dimensionless curvature leads to a lower value of $KC_D$ where vortex shedding is initiated. Consequently, vortex shedding at the sharp edged big opening will take place at a lower value of the \textit{local} Keulegan--Carpenter number than the vortex shedding from the jet pump waist.

\section{Conclusions}
\label{sec:conclusions}

A CFD based parameter study is performed to investigate the influence of various geometric parameters on the performance of jet pumps for thermoacoustic applications. A total of 19 different jet pump geometries are used and for each geometry a number of wave amplitudes are simulated, resulting in a total of 197 simulations. In correspondence with previous work, four flow regimes are distinguished and separated in a fixed variable space.\cite{Oosterhuis2015} This space spans the jet pump taper angle $\alpha$, the dimensionless curvature $\chi$ and the Keulegan--Carpenter number $KC_D$. At a certain value of $KC_D$, single-sided vortex propagation from the jet pump waist is initiated and the flow field becomes asymmetric between the left and right side of the jet pump. This onset is found at $KC_D=0.5$ for a dimensionless curvature of $\chi=0.36$. Two different expressions are proposed to account for the effect of $\chi$, but more data is required to decisively choose one of the two. At some point, flow separation inside the jet pump can be distinguished and vortices are shed from the jet pump waist in alternating directions leading to a flow field that is more symmetric. This transition is found to be strongly dependent on the jet pump taper angle and occurs when $KC_D > \frac{1}{\alpha}\left(1+0.15/\chi\right)^{-1}$.

The amount of asymmetry in the flow field has a direct consequence on the measured time-averaged pressure drop. When no vortex shedding occurs, no dimensionless time-averaged pressure drop is observed. Furthermore, as soon as the flow separation is observed, the dimensionless pressure drop has already decayed significantly compared to the maximum value. For all the performed simulations, the maximum jet pump effectiveness is observed when either single-sided or two-sided vortex propagation occurs. Consequently, the practical operation area of jet pumps with $D_s/\delta_\nu \gg 1$ and $Re<Re_c$ is bound between the onset of vortex shedding from the jet pump waist and the occurrence of flow separation. This can be used as a guideline for future jet pump design.

The design considerations presented in this paper, are supposed to provide better insight into the validity of the quasi--steady approximation that is widely used for the design of jet pumps for thermoacoustic applications. The quasi--steady approximation is in most cases an ideal representation of the jet pump performance and is valid only in a small operation area. Outside this region, the actual jet pump performance is expected to be significantly lower and a correction using the presented results is advised.

Although the current parameter study represents a wide range of jet pump geometries, it is restricted to jet pumps with a single, linear tapered hole. As such, this work can serve as a basis for further jet pump geometry optimization in terms of effectiveness, compactness and robustness. Extensions to multiple holes or different shaped jet pump walls might affect the jet pump performance and the occurrence of flow separation. Furthermore, the influence of turbulence on the flow separation requires further attention. If the flow separation regime could be shifted to higher Keulegan--Carpenter numbers without a decrease in effectiveness, the operation area of a jet pump can be extended, greatly enhancing its robustness.

\medskip

\noindent \textbf{Acknowledgements}

\setlength{\parindent}{0.7cm} 
The authors would like to gratefully thank Bosch Thermotechnology and Agentschap~NL for the financial support as part of the EOS-KTO research program under project number KTOT03009.

\end{document}